\documentclass[10pt, conference, compsocconf]{IEEEtran}
\usepackage{graphicx}
\usepackage{caption}
\usepackage{subcaption}
\usepackage[ruled,vlined]{algorithm2e}
\usepackage{soul}
\usepackage{comment}
\usepackage{nth}
\usepackage{multirow}
\usepackage{makecell}
\usepackage{enumitem}
\usepackage{microtype}
\usepackage{cite}
\usepackage{amsmath,amssymb,amsfonts}
\usepackage{algorithmic}
\usepackage{textcomp}
\usepackage{xcolor}
\usepackage{hyperref}
\hypersetup{
	colorlinks,
	citecolor=black,
	filecolor=black,
	linkcolor=black,
	urlcolor=black,
	linktoc=all,
	pageanchor=false
}

\usepackage[capitalise,noabbrev]{cleveref}
\usepackage{url}
\def\BibTeX{{\rm B\kern-.05em{\sc i\kern-.025em b}\kern-.08em
    T\kern-.1667em\lower.7ex\hbox{E}\kern-.125emX}}

\usepackage{tikz}

\newcommand*\annotatedFigureText[4]{\node[draw=none, anchor=south west, text=#2, inner sep=0, text width=#3\linewidth,font=\small\sffamily] at (#1){#4};}
\newenvironment {annotatedFigure}[1]{\centering\begin{tikzpicture}
\node[anchor=south west,inner sep=0] (image) at (0,0) { #1};\begin{scope}[x={(image.south east)},y={(image.north west)}]}{\end{scope}\end{tikzpicture}}

\newcommand\blfootnote[1]{%
  \begingroup
  \renewcommand\thefootnote{}\footnote{#1}%
  \addtocounter{footnote}{-1}%
  \endgroup
}
\IEEEoverridecommandlockouts
\def\BibTeX{{\rm B\kern-.05em{\sc i\kern-.025em b}\kern-.08em
    T\kern-.1667em\lower.7ex\hbox{E}\kern-.125emX}}

\begin{document}

\title{hEARt: Motion-resilient Heart Rate Monitoring \\ with In-ear Microphones}

\author{\IEEEauthorblockN{Kayla-Jade Butkow\IEEEauthorrefmark{1}, Ting Dang\IEEEauthorrefmark{1}, Andrea Ferlini\IEEEauthorrefmark{1}, Dong Ma\IEEEauthorrefmark{2} and Cecilia Mascolo\IEEEauthorrefmark{1}} \IEEEauthorblockA{\IEEEauthorrefmark{1}University of Cambridge, Cambridge, UK} \IEEEauthorblockA{\IEEEauthorrefmark{2}Singapore Management University, Singapore\\
\{kjb85, td464, af679\}@cam.ac.uk, dongma@smu.edu.sg, cm542@cam.ac.uk}
}

\maketitle
\thispagestyle{plain}
\pagestyle{plain}
\begin{abstract}
With the soaring adoption of in-ear wearables, the research community has started investigating suitable in-ear heart rate (HR) detection systems.
HR is a key physiological marker of cardiovascular health and physical fitness. Continuous and reliable HR monitoring with wearable devices has therefore gained increasing attention in recent years. Existing HR detection systems in wearables mainly rely on photoplethysmography (PPG) sensors, however, these are notorious for poor performance in the presence of human motion.  
In this work, leveraging the occlusion effect that enhances low-frequency bone-conducted sounds in the ear canal,  we investigate for the first time \textit{in-ear audio-based motion-resilient} HR monitoring. We first collected HR-induced sounds in the ear canal leveraging an in-ear microphone under stationary and three different activities (i.e., walking, running, and speaking). Then, we devised a novel deep learning based motion artefact (MA) mitigation framework to denoise the in-ear audio signals, followed by an HR estimation algorithm to extract HR.  With data collected from 20 subjects over four activities, we demonstrate that hEARt, our end-to-end approach, achieves a mean absolute error (MAE) of 3.02 $\pm$ 2.97~BPM, 8.12 $\pm$ 6.74~BPM, 11.23 $\pm$ 9.20~BPM and 9.39 $\pm$ 6.97~BPM for stationary, walking, running and speaking, respectively, opening the door to a new non-invasive and affordable HR monitoring with usable performance for daily activities.
Not only does hEARt outperform previous in-ear HR monitoring work, but it outperforms reported in-ear PPG performance.
\end{abstract}
\blfootnote{This work is supported by ERC through Project 833296 (EAR), and Nokia Bell Labs through their donation for the Centre of Mobile, Wearable Systems and Augmented Intelligence to the University of Cambridge.}


\vspace{-0.5em}

\section{Introduction}
Heart rate (HR) is an excellent indicator of fitness level, and is strongly associated with cardiovascular disease and mortality risk. HR monitoring can help design workout routines to maximize training effect, and, more importantly, serves as an early biomarker for heart disease since cardiovascular fitness is a key predictor of cardiovascular disease. Additionally, heart rate variability (HRV), the change in time between successive beats, is a predictor of physical and mental health. HRV, a proxy for autonomic nervous system behaviour, is predictive of aerobic fitness when measured during both maximal and sub-maximal exercise~\cite{CardiacAutonomicResponsesDuringExerciseAndPostexercise}. Thus measuring HR under motion is critical for monitoring human health and wellbeing.% \td{remove HRV sentence here as we don't focus on this?} \kb{This was there to epmhasise why recreating ECG is important as opposed to just using DL to predict HR} 

Electrocardiographic (ECG) telemetry monitoring is the standard for HR and HRV monitoring, however ECGs need to be connected to the body with leads making them unsuitable for realistic and mobile settings. Although attempts to devise portable ECG, such as ECG chest straps, have been introduced, they remain cumbersome, uncomfortable, and inconvenient. New smartwatches include a single-lead ECG, however they require the user to remain still and to close the ECG circuit with their fingers. They are thus unable to monitor continuously.

Recent trends in wearables have led to a proliferation of studies investigating different sensors on smartwatches, earables, and other wearables for HR monitoring. Photoplethysmography (PPG) sensors, which measure light scatter as a result of blood flow, are most commonly adopted due to their non-invasiveness, easy implementation and low cost. Although PPG is effective and accurate for HR measurements under stationary conditions~\cite{Bent2020InvestigatingSensors}, it is sensitive to motion artefacts (MAs) caused by user's body movement or physical activities~\cite{navalta2020concurrent, Bent2020InvestigatingSensors, ahn2020device}. Due to these MAs, the research community has yet to find an agreement on the goodness of wrist-worn PPG (e.g. PPG on smartwatch).
While the topic has been widely investigated~\cite{ahn2020device, Bent2020InvestigatingSensors, navalta2020concurrent}, a consensus on the best commercially available device to monitor the wearer’s HR whenever motion is concerned, is yet to be found. 
Moreover, intense motion, like walking and running, yields substantial deviations from ground-truth (GT), resulting in average errors up to 30\% across a wide-spectrum of wrist-worn devices~\cite{Bent2020InvestigatingSensors}. Dealing with interference from MAs is thus an open and challenging problem in HR estimation.

% Other factors may come into play as well:
% the fit of the watch is key to achieve acceptable HR measurements. A more comfortable/loose fit of the smartwatch correlates with drastic performance degradation~\cite{ahn2020device};
% skin perfusion and the presence of tattoos are reported to be a deterrent to accurate HR monitoring~\cite{Bent2020InvestigatingSensors}.

Due to the limitations of wrist-based PPG, researchers have started investigating alternative wearables for HR monitoring under motion. With the rapid spreading of ear-worn wearables (earables) in daily life~\cite{EarablesForPersonalScaleBehaviorAnalytics}, earables can be a portable and non-invasive means of continuous HR detection. Particularly, due to their pervasiveness during physical activity (specifically while walking and running), the earable form factor can be exploited for HR monitoring while under motion. Research has started to emerge in earable based PPG for continuous HR sensing~\cite{goverdovsky2017hearables}. However, despite being a promising modality, real world performance of earable PPG under motion is still poor~\cite{navalta2020concurrent,ferlini2021ear}.
Indeed, similar to what is observed for wrist-worn devices~\cite{Bent2020InvestigatingSensors}, errors around 30\% have been reported~\cite{ferlini2021ear}.

Current commercial earables are equipped with multiple sensors, including outer and inner ear microphones which fulfil fundamental functionalities of the device (e.g., speech detection and active noise cancellation).
Recently, Martin and Voix~\cite{InEarAudioWearableMeasurementOfHeartAnd} proposed to measure HR using a microphone placed in the human ear canal. When the ear canal opening is sealed by the earbuds, the cavity formed between the ear tip and eardrum enables an enhancement of low-frequency sounds, called the occlusion effect~\cite{Stone2014AHz}. As a result, heartbeat-induced sounds that propagate to the ear canal through bone conduction are amplified and can be leveraged for HR estimation. Their results show an error of 5.6\% for HR determination under stationary conditions. However, \cite{InEarAudioWearableMeasurementOfHeartAnd} only demonstrated the feasibility of measuring HR with in-ear microphones while an individual is stationary: {\em how in-ear microphone HR measurement performs under active scenarios remains unclear and unexplored}.

% Like with in-ear PPG sensing, in-ear audio sensing for HR monitoring is also a recent area of research, with very scarce literature.
% Recent work~\cite{InEarAudioWearableMeasurementOfHeartAnd} has proposed in-ear audio using a microphone for HR and respiratory rate measurements. 
% Their results show a MAE of 4.3~BPM and a mean difference estimate of –0.44~BPM with a limit of agreement (LoA) interval of –14.3
% to 13.4~BPM.
% This points at the potential of in-ear audio-based HR monitoring. 
% However, the work only investigates the stationary setting, and a higher error is reported for the time period when the subject’s body is in motion.
% Motivated by these findings, we aim to explore the power of in-ear audio sensors for accurate HR detection under typical daily motion conditions including walking, running, and speaking. 

In this work, we focus on in-ear HR estimation under both stationary and active scenarios (e.g., walking, running and speaking). The biggest hurdle to accurate HR measurement is motion-induced interference, which is amplified by the occlusion effect along with the heart sounds~\cite{Ma2021OESense}.
%originating from the fact that the occlusion effect amplifies not just the heartbeat-induced sounds, but also the sounds/vibrations generated by these activities~\cite{Ma2021OESense}. 
Removing such interference is non-trivial and poses three challenges. First, the strength of heartbeats is much weaker than the foot strikes, so heartbeat signals are buried in the walking signals. Second, since HR and walking frequency (i.e. cadence) are similar (both around 1.5-2.3~Hz~\cite{murray1985treadmill}), it is hard to separate them in the frequency domain. Third, due to the proximity of the ear to the human vocal system, human speech and its associated jaw movements can overwhelm the heartbeat-induced sound. 

To address these challenges, we propose a processing pipeline for accurate HR detection in the presence of MAs, namely, walking, running and speaking. Different from previous audio-based HR estimation works~\cite{passler2019ear, patterson2009flexible, goverdovsky2017hearables, InEarAudioWearableMeasurementOfHeartAnd}, we also validate the functioning of our technique in the presence of speech, showing how the proposed approach can successfully deal with speaking activities.
%\kb{Is the rest of this section needed given that we summarise the contributions at the end of the chapter?} Specifically, (1) based on the challenges in separating heart and activity sounds, we develop a convolutional neural network (CNN) to clean heart sounds (HS) with the supervision of ground truth (GT) ECG signals.
% ECG signals are only required during model training, but are not needed for testing or deployment. We further leverage transfer learning that pre-trains the denoising CNN using a large HS database (PASCAL) and fine-tunes it using our data to account for the small data set. (2) using the resultant cleaned signals, we compute the HR using a simple signal processing pipeline.
% (3) differently from previous audio-based HR estimation works~\cite{passler2019ear, patterson2009flexible, goverdovsky2017hearables, InEarAudioWearableMeasurementOfHeartAnd}, we additionally validate the functioning of our technique in the presence of human speech, showing how the proposed approach can successfully deal with speaking activities.
With data collected from 20 subjects, we demonstrate that an in-ear microphone can be a viable sensor for HR estimation under motion cases with good performance. Specifically, with mean absolute percentage error (MAPE) less than 10\% while stationary, walking and running, this system is accurate according to ANSI standards for HR accuracy for a physical monitoring device~\cite{PhysicalActivityMonitoringForHeartRateANSI,Bent2020InvestigatingSensors}. Additionally, 
% As a coarse comparison, such performance is even superior to most of the PPG-based approaches as reported in the literature~\cite{zhang2019motion, Bent2020InvestigatingSensors, ferlini2021ear}. \kb{Maybe we want to rather not say this and stick to a direct ear based ppg comparison?}
%Specifically, 
because of the artifacts considered, the vantage points (the ears), and the device form-factor (earables), our work is directly comparable to~\cite{ferlini2021ear}.
Notably, we significantly outperform in-ear PPG~\cite{ferlini2021ear} (65\% and 67\% improvement) for walking and running. This result hints at the great potential of in-ear microphones for cardiovascular health monitoring, even under challenging scenarios.
Moreover, compared to PPG, microphones are more energy efficient~\cite{datasheet:SPU1410LR5H-QB,datasheet:MAXM86161} and affordable offering additional appeal for continuous HR estimation. 

% Notably, we significantly outperform \cite{ferlini2021ear} in walking and running, and marginally for talking, achieving a MAPE of 9.53\%, 9.80\% and 12.06\% respectively. This is 65\% and 67\% better than in-ear PPG for walking and running.
% This result hints at the great potential of in-ear microphones for cardiovascular health monitoring, even under challenging scenarios.
% Moreover, compared to PPG, microphones are more energy efficient~\cite{datasheet:SPU1410LR5H-QB,datasheet:MAXM86161} and affordable offering additional appeal for continuous HR estimation. 

% \cm{we could cut the contributon list}
The contribution of this work can be summarized as follows: \textbf{(i)} We explore HR estimation with in-ear microphones and present an analysis of the interference imposed by common human activities. \textbf{(ii)} We propose a novel pipeline for HR estimation under MAs,  consisting of a CNN-based module using U-Net architecture to enhance audio-based heart sounds (HS) with ECG as a reference, and an estimation module using signal processing to estimate HR from cleaned signals. We further leverage transfer learning that pre-trains the model using a large HS dataset and fine-tunes it using our data to effectively capture HS related information, and handle the limited data size. To the best of our knowledge, no previous works have attempted to clean and enhance audio-based HS captured by earables using ECG signals. \textbf{(iii)} We built an earbud prototype with good signal-to-noise ratio (SNR) and collected data from 20 subjects%, which we will release to the research community
. \textbf{(iv)} Results show that we can achieve mean absolute errors of 3.02 $\pm$ 2.97~BPM, 8.12 $\pm$ 6.74~BPM, 11.23 $\pm$ 9.20~BPM and 9.39 $\pm$ 6.97~BPM for stationary, walking, running and speaking, respectively, demonstrating the effectiveness of the proposed approach in combating MAs.

\section{Primer}
\label{sec:primer}
% In this section, we motivate our work by providing a brief evolution of HR detection techniques, including the preliminary research into in-ear microphone based HR detection in a static setting. We then present the challenges of achieving accurate and portable in-ear HR estimation in more realistic conditions.

In this section, we present the mechanism by which HS are collected in the ear and the challenges of achieving accurate and portable in-ear HR estimation under motion conditions.

\subsection{In-ear Heart Sound Acquisition}

\begin{figure}[ht]
	\centering
	\includegraphics[width=0.46\linewidth]{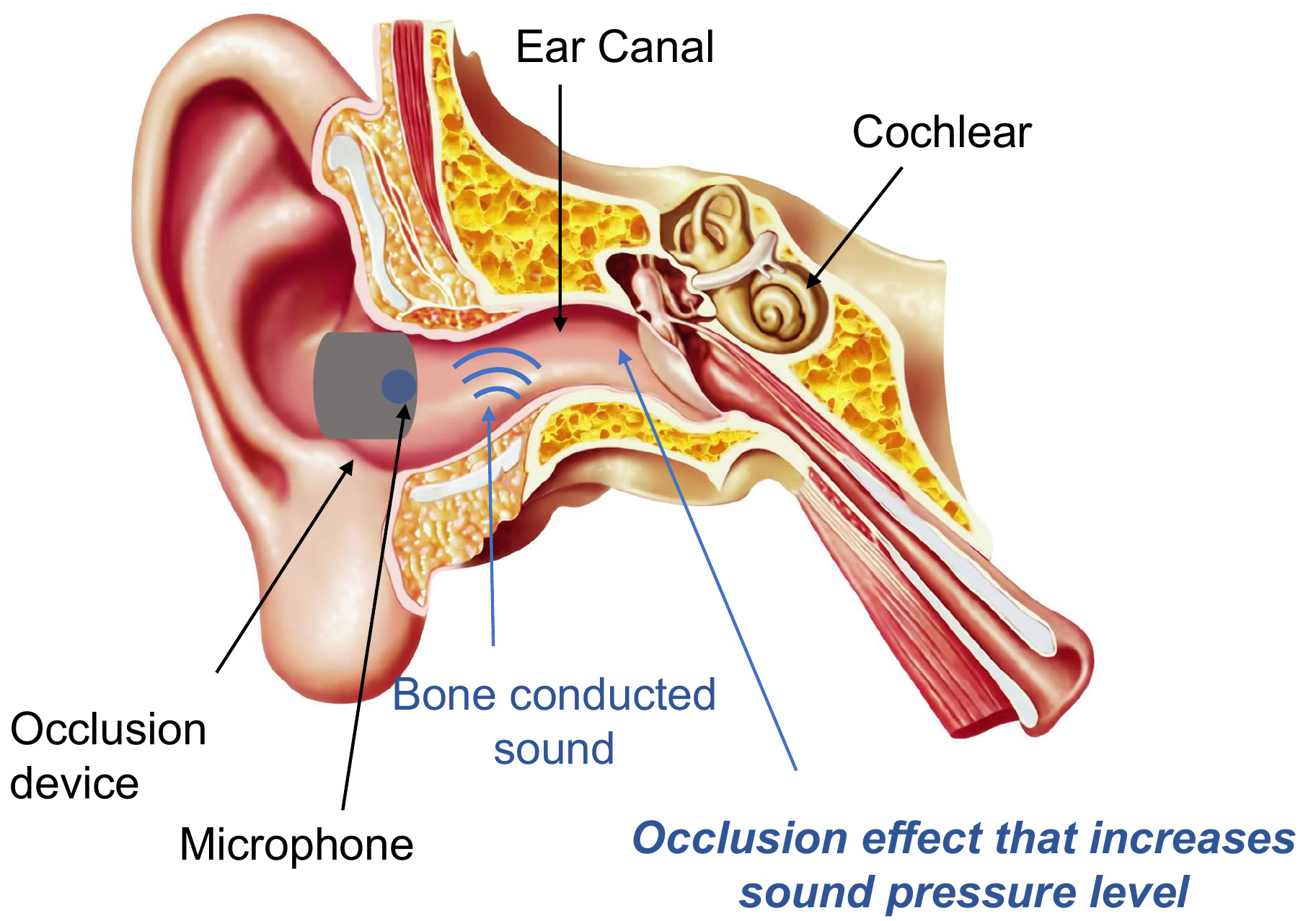}%occlusion_effect
	\caption{The occlusion effect and the anatomy of the ear.}
	\label{fig:occlusioneffect}
\end{figure}

\begin{figure}[ht]
\centering
\begin{subfigure}[t]{.48\columnwidth}
  \begin{annotatedFigure}
	{\includegraphics[width=.9\linewidth]{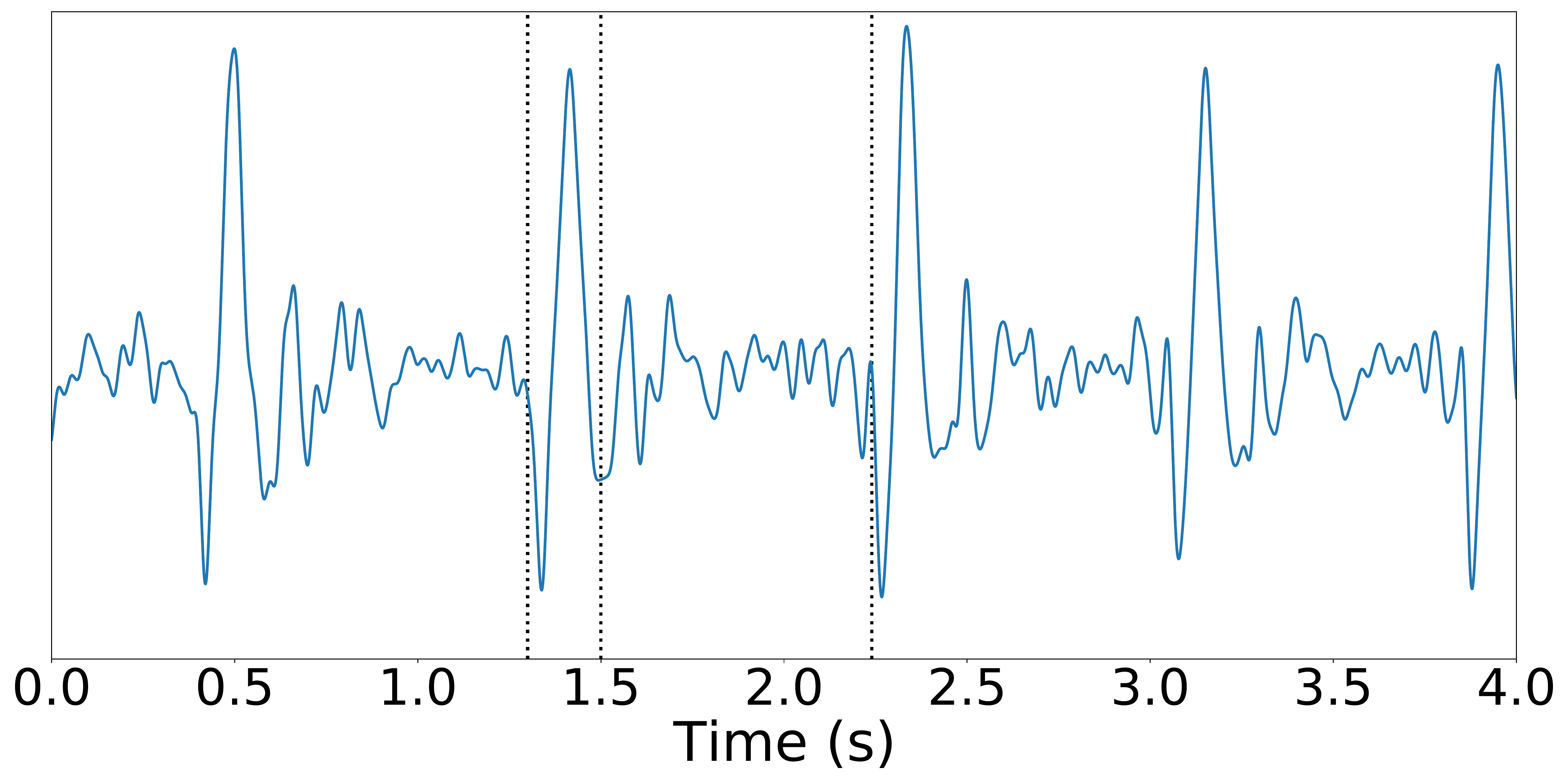}}
	\annotatedFigureText{0.4,0.2515}{black}{0.01}{S2}
	\annotatedFigureText{0.369,0.802}{black}{0.01}{S1}
\end{annotatedFigure}
\end{subfigure}
\begin{subfigure}[t]{.48\columnwidth}
\begin{annotatedFigure}
	{\includegraphics[width=0.9\linewidth]{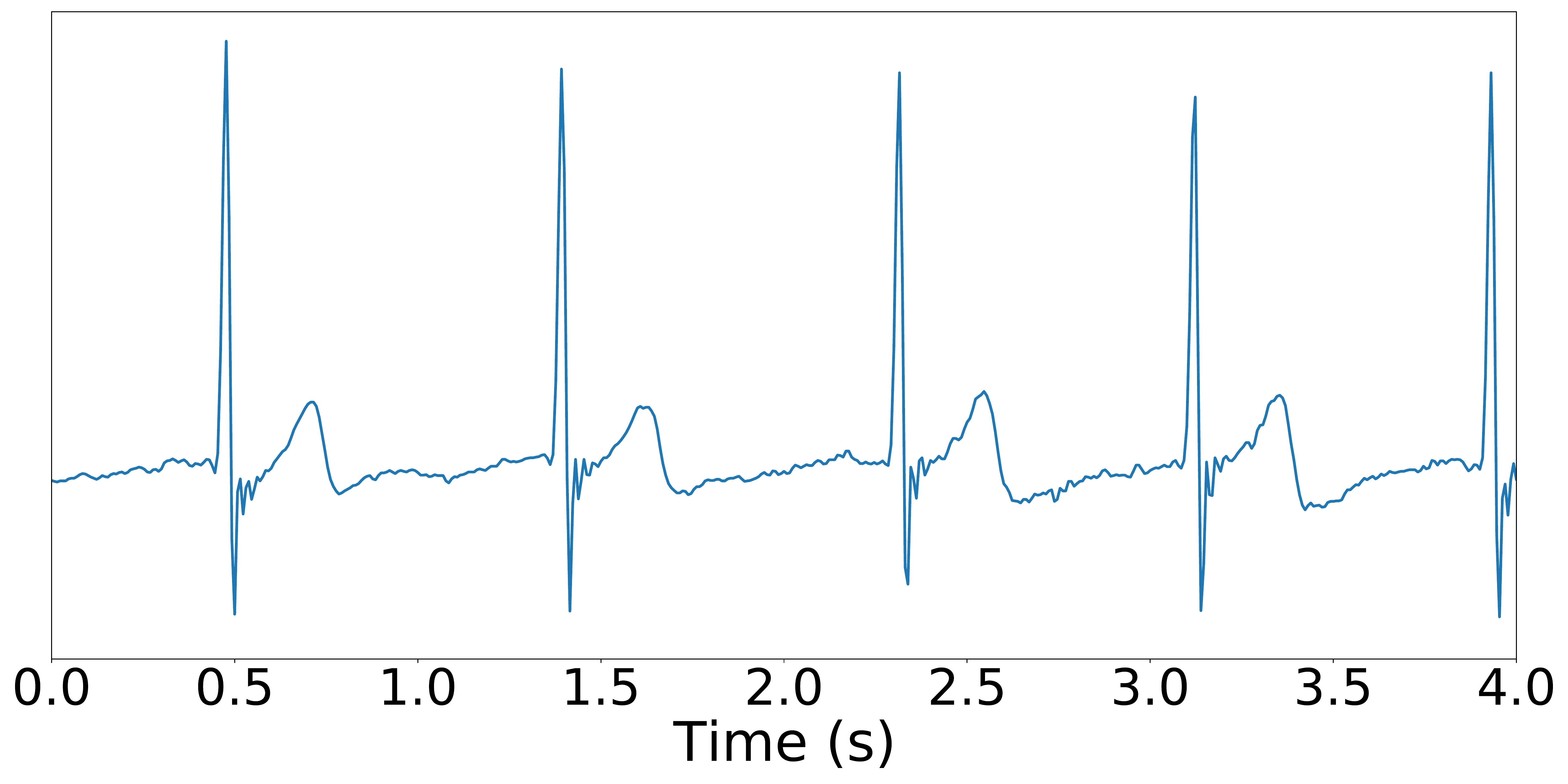}}
	\annotatedFigureText{0.395,0.4895}{black}{0.04}{T}
	\annotatedFigureText{0.369,0.758}{black}{0.04}{R}
	\annotatedFigureText{0.512,0.434}{black}{0.04}{P}
	\annotatedFigureText{0.28,0.2}{black}{0.04}{Q}
	\annotatedFigureText{0.373,0.2}{black}{0.04}{S}
\end{annotatedFigure}
\end{subfigure}
% \vspace*{-5mm}
\caption{The (left) sound signal captured by the internal microphone and the (right) corresponding ECG signal.}
\label{fig:sound_vs_ecg}
\end{figure}

%We now describe the mechanism by which heart sounds are generated and captured inside the ear canal.\td{remove this sentence to save space?} 
Bone conduction, a physiological phenomenon whereby sound is conducted through the bones directly to the inner ear, causes vibrations in the walls of the ear~\cite{Stone2014AHz}. When the ear canal is occluded, the increase in impedance at the entrance of the ear canal results in an amplification of low frequency sounds conducted by the bones~\cite{Stone2014AHz}. This effect, illustrated in \cref{fig:occlusioneffect}, is known as the occlusion effect. Since bone conveys low-frequency sounds~\cite{tonndorf1968new}, the bone-conducted HS are amplified in the occluded ear canal~\cite{InEarAudioWearableMeasurementOfHeartAnd}. Heart sounds can thus be detected using a microphone placed inside the occluded ear canal. An example showing the HS captured by the internal microphone is shown in \cref{fig:sound_vs_ecg}. Clearly, the two sounds in the cardiac cycle (S1 and S2) can be captured using the in-ear microphone, thus indicating the potential of in-ear microphones for HR monitoring. The correlation between the in-ear captured audio and the ECG signal is also evident in \cref{fig:sound_vs_ecg}.
%Here, the QRS complex (the combination of three of the graphical deflections seen on a typical electrocardiogram\footnote{\url{https://en.wikipedia.org/wiki/QRS\_complex}}) of the ECG, as shown in \cref{fig:sound_vs_ecg}, which occurs due to ventricular depolarization~\cite{AssessmentOfCardiovascularFunction}, corresponds to the S1 heart sound.
%The T and P waves seen in \cref{fig:sound_vs_ecg} occur due to ventricular repolarization and atrial depolarization respectively. The end of the ECG T wave corresponds to the S2 heart sound~\cite{AssessmentOfCardiovascularFunction}.\td{maybe remove the description of qrs, t p wave, and just 'they are similar' is enough?}

\begin{figure}[t]
	\centering
	\begin{subfigure}{.24\columnwidth}
		\centering
		% include fourth image
		\includegraphics[width=1\linewidth]{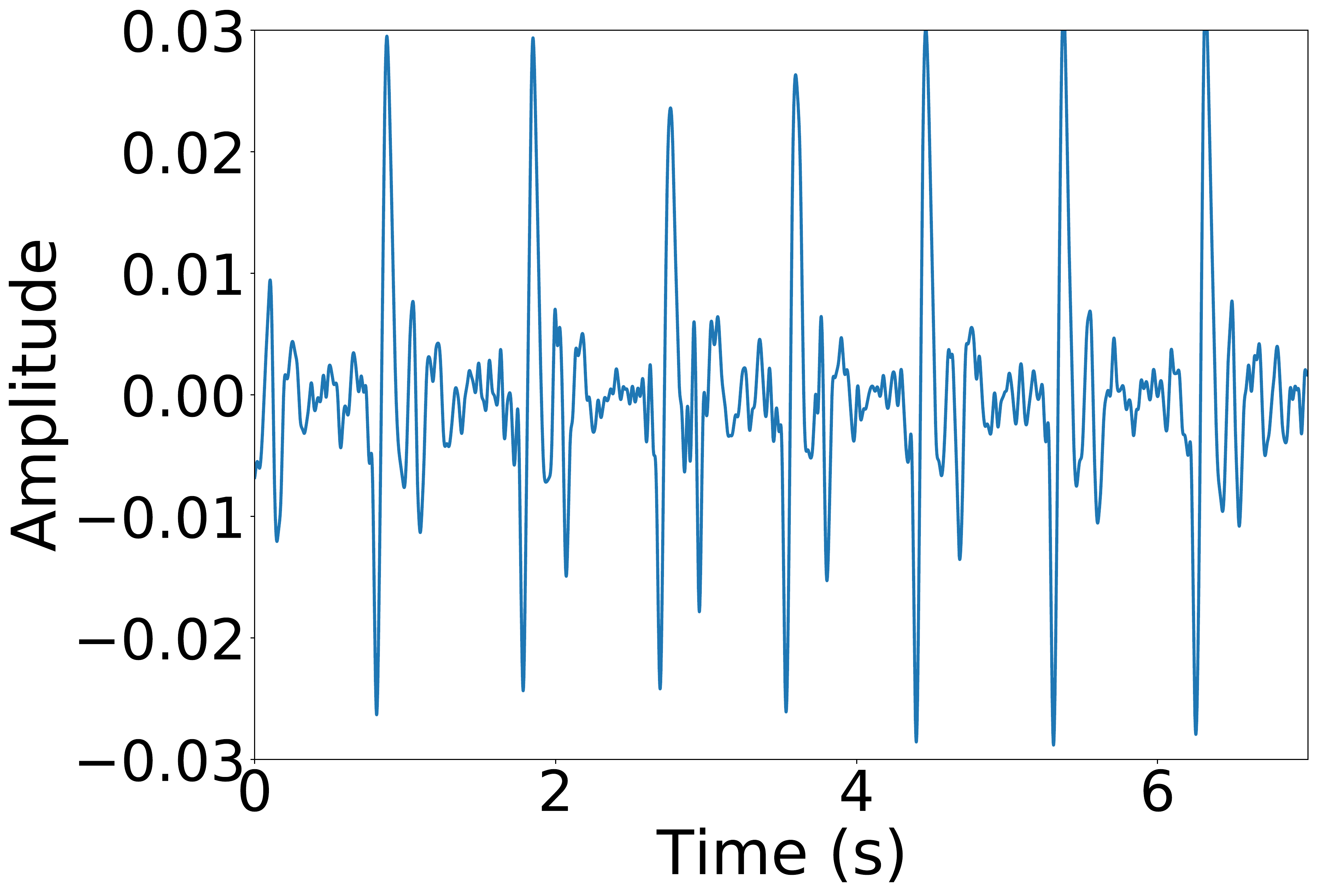}  
		\caption{Stationary}
		\label{fig:raw_sig_still}
	\end{subfigure}
	\begin{subfigure}{.24\columnwidth}
		\centering
		% include first image
		\includegraphics[width=1\linewidth]{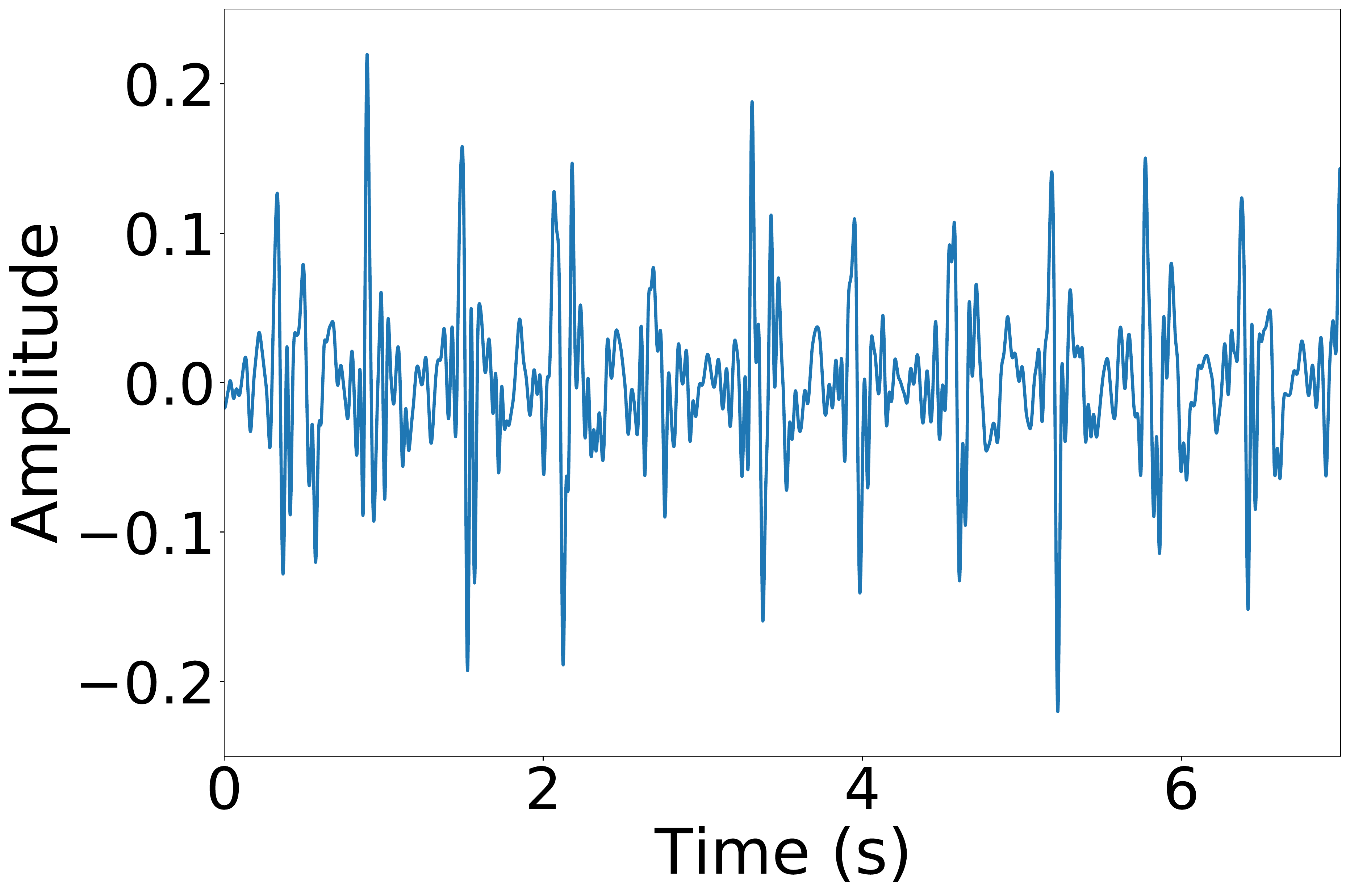}  
		\caption{Walking}
		\label{fig:raw_sig_walk}
	\end{subfigure}
	\begin{subfigure}{.24\columnwidth}
		\centering
		% include second image
		\includegraphics[width=1\linewidth]{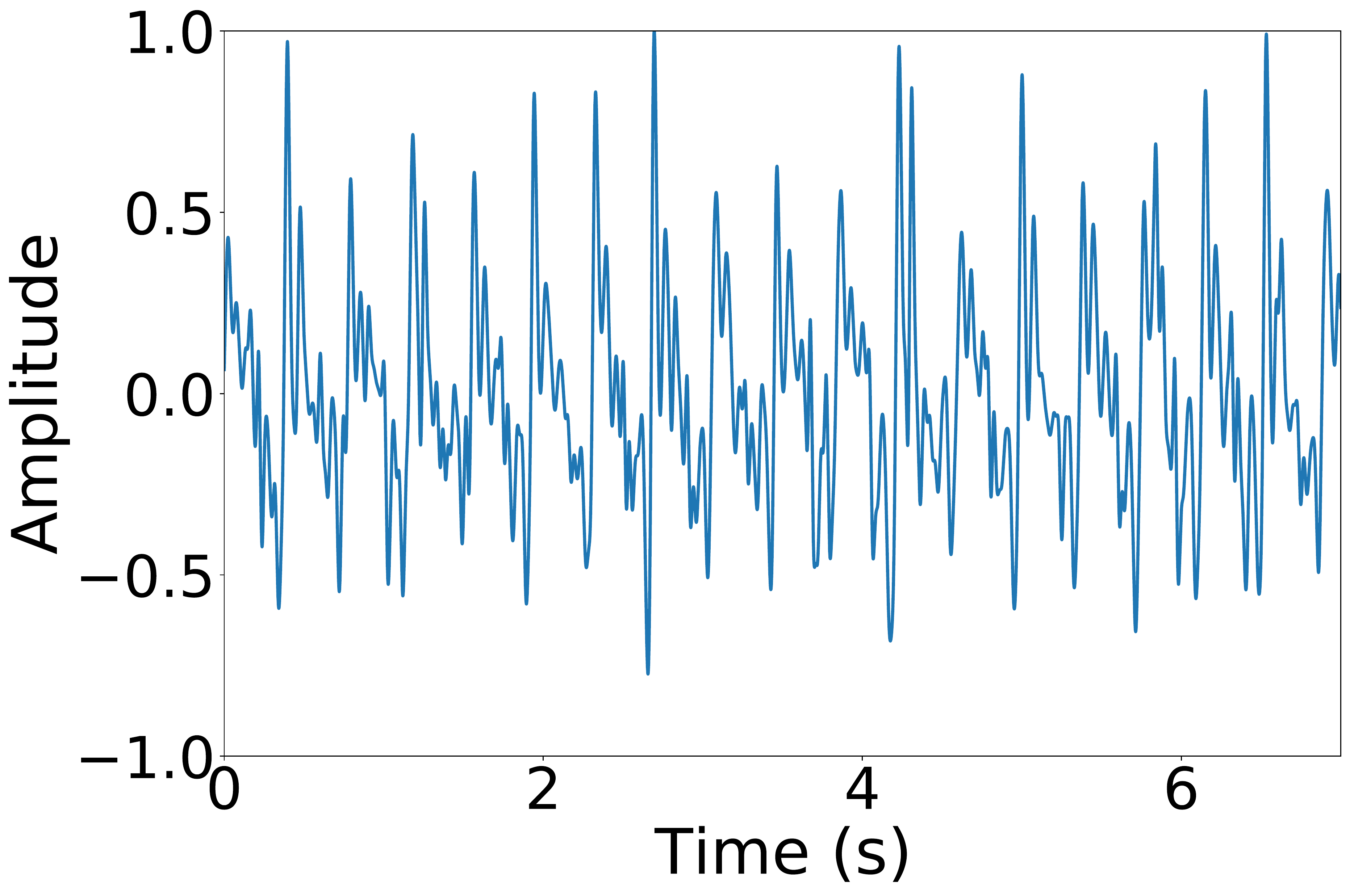}  
		\caption{Running}
		\label{fig:raw_sig_run}
	\end{subfigure}
		\begin{subfigure}{.24\columnwidth}
		\centering
		% include second image
		\includegraphics[width=1\linewidth]{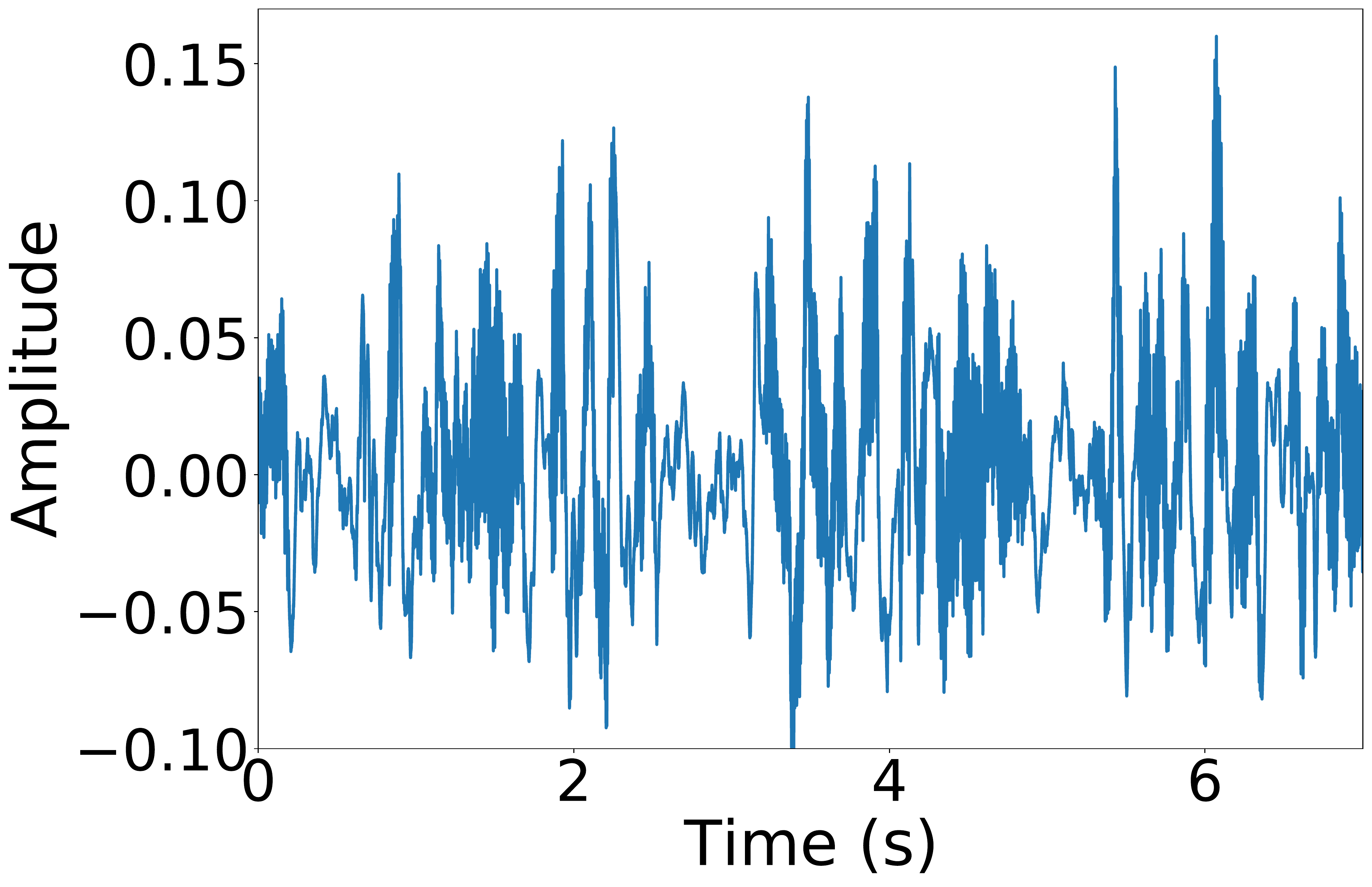}  
		\caption{Speaking}
		\label{fig:raw_sig_speak}
	\end{subfigure}
   	\begin{subfigure}{.24\columnwidth}
		\centering
		% include fourth image
		\includegraphics[width=1\linewidth]{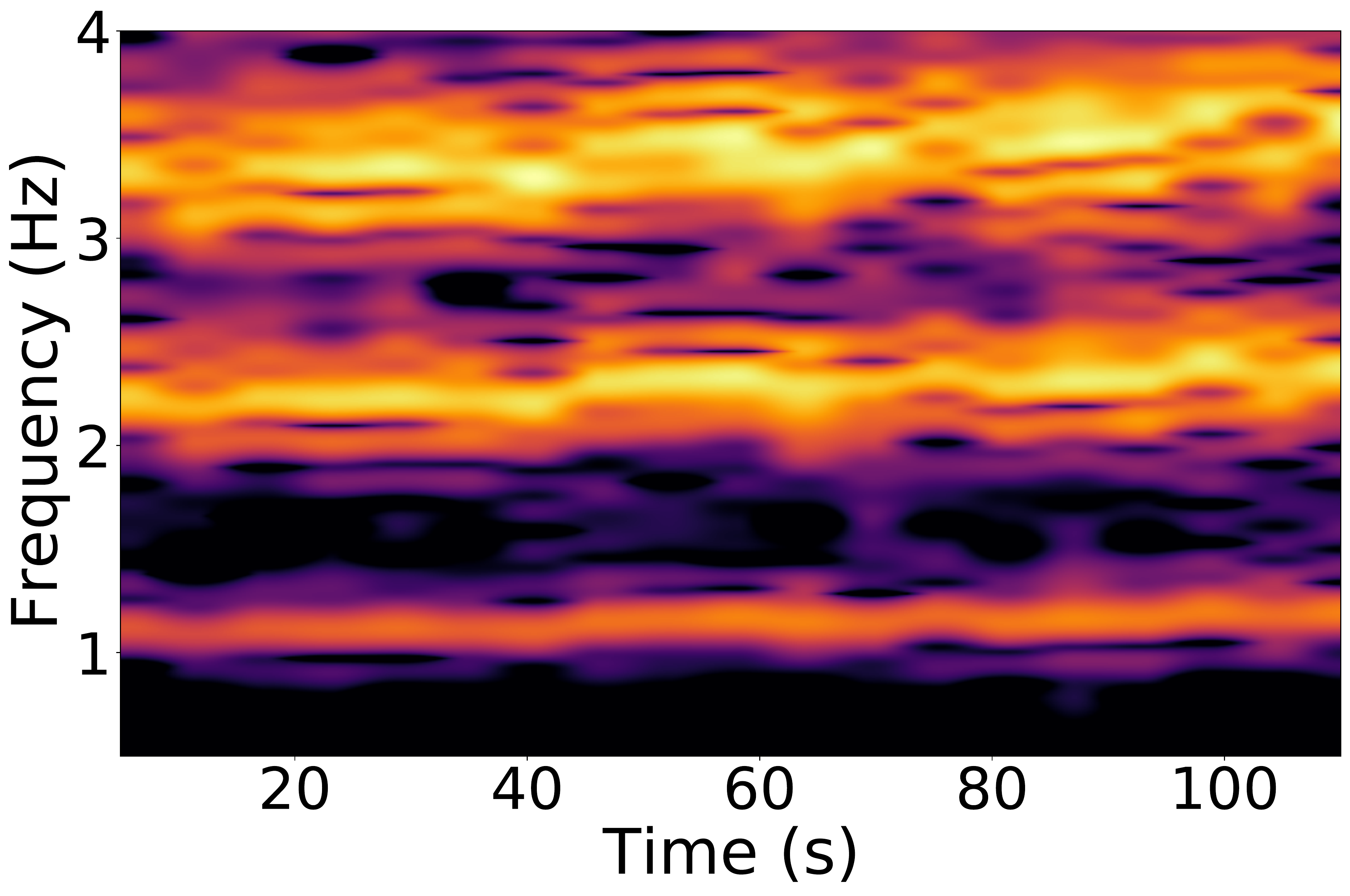}  
		\caption{Stationary}
		\label{fig:still_spec}
	\end{subfigure}
 	\begin{subfigure}{.24\columnwidth}
		\centering
		% include first image
		\includegraphics[width=1\linewidth]{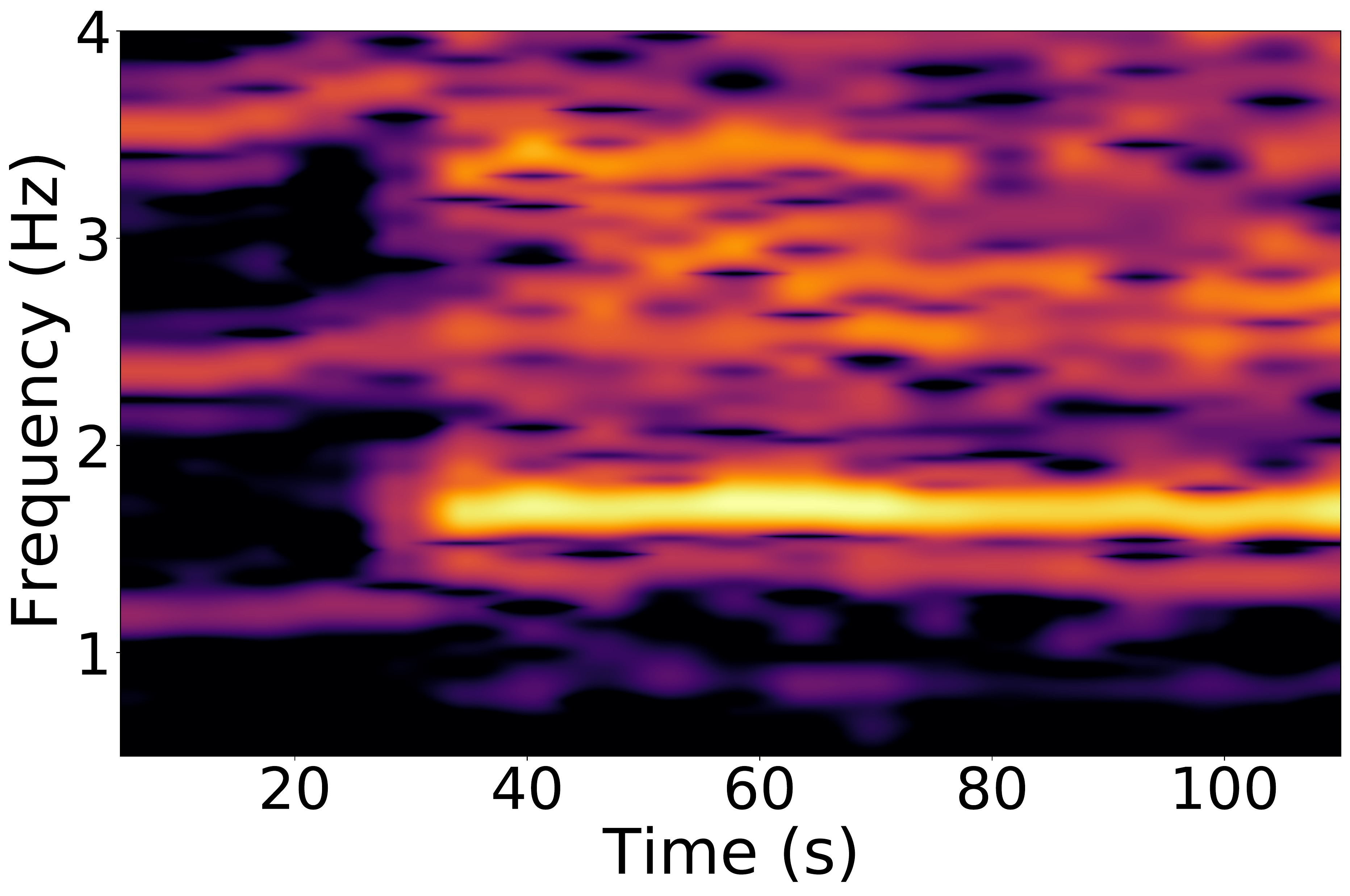}  
		\caption{Walking}
		\label{fig:walking_spec}
	\end{subfigure}
	\begin{subfigure}{.24\columnwidth}
		\centering
		% include second image
		\includegraphics[width=1\linewidth]{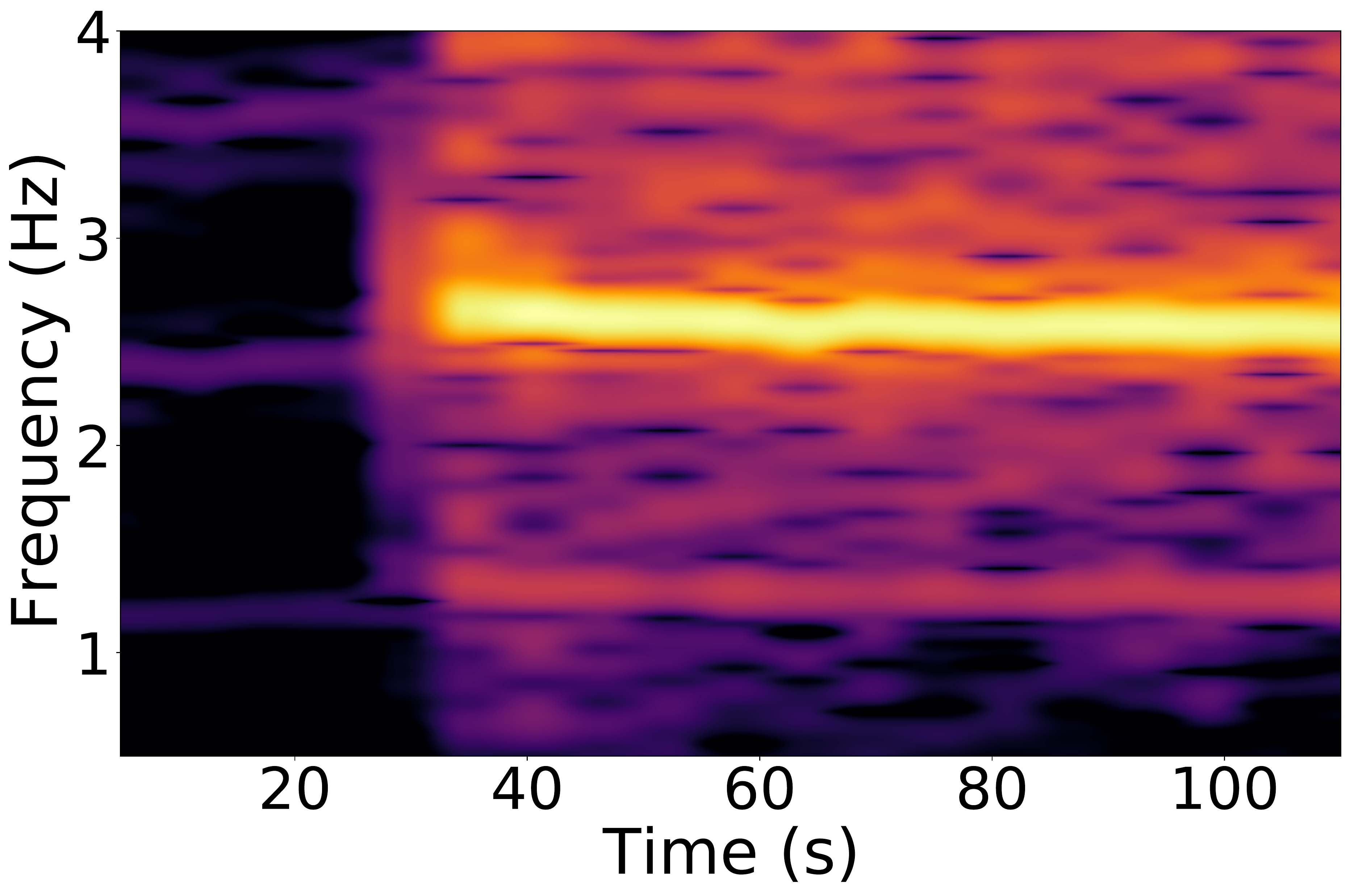}  
		\caption{Running}
		\label{fig:running_spec}
	\end{subfigure}
		\begin{subfigure}{.24\columnwidth}
		\centering
		% include second image
		\includegraphics[width=1\linewidth]{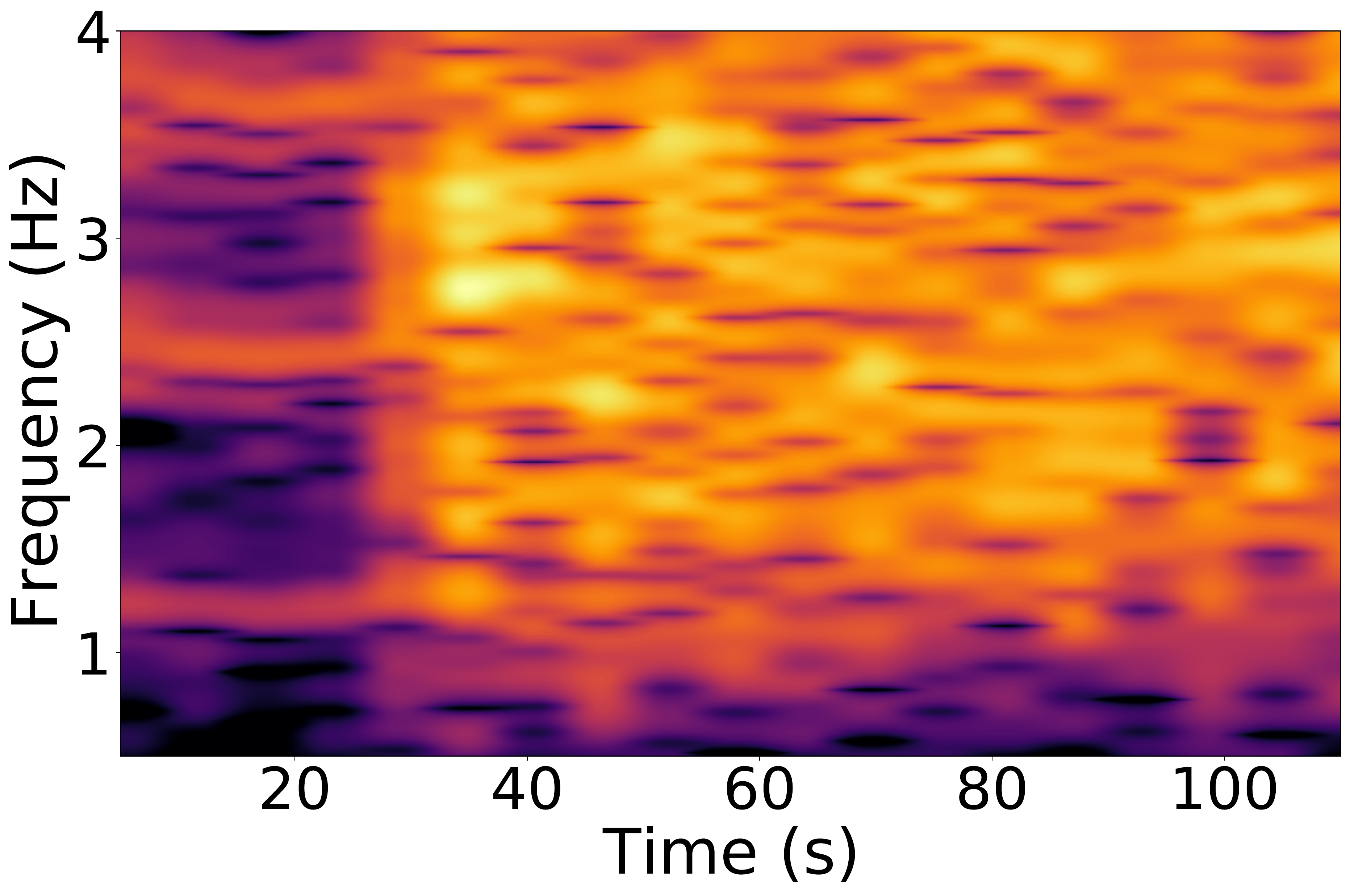}  
		\caption{Speaking}
		\label{fig:talking_spec}
	\end{subfigure}
	\caption{Time domain representations and spectrograms of audio signals captured by the in-ear microphone.}
	\label{fig:long_spec}
\end{figure}

\subsection{Motion Artefacts Analysis and Challenges}
\label{sec:challenges}
In-ear microphone based HR estimation suffers from human MAs since the occlusion effect not only amplifies the heartbeat-induced sound, but also enhances other bone-conducted sounds and vibrations inside the body~\cite{ferlini2021eargate,Ma2021OESense}. \cref{fig:raw_sig_still,fig:raw_sig_walk,fig:raw_sig_run,fig:raw_sig_speak} illustrates the recorded audio signals from the in-ear microphone while stationary, walking, running and speaking within a seven second window. \cref{fig:still_spec,fig:walking_spec,fig:running_spec,fig:talking_spec} are spectrograms of the activities shown over a longer timescale so that trends in HR can be seen. The heartbeat is clearly observable when an individual is stationary in \cref{fig:raw_sig_still}, with frequency lying around 1.2Hz and its \nth{1} and \nth{2} harmonics clearly observable in \cref{fig:still_spec}. Contrastingly, it is completely overwhelmed by the amplified step sounds in \cref{fig:raw_sig_walk} (note the different scales of the y-axis), with the periodic peaks corresponding to the sound of foot strikes that propagate through the human skeleton, resulting in a significantly higher energy observed around 1.7Hz (the cadence) in \cref{fig:walking_spec}. Though the HR and its harmonics are still observable, it is difficult to estimate HR directly from the raw corrupted audio signals. Furthermore, it is evident that the periods of HS and walking are similar, resulting in an overlap in the frequency domain, making it challenging to split the HS and walking signals and estimate the HR either in time domain or frequency domain. 

The heartbeats are further affected by foot strikes during running (\cref{fig:raw_sig_run}) that exhibit far stronger energy than any of the other activities, with high energy in 2.6Hz (\cref{fig:running_spec}) corresponding again to the cadence. The speech sound in \cref{fig:raw_sig_speak} also shows strong noise amplitudes due to the proximity of the ear and human vocal system, making the heartbeat-induced sound indiscernible. 
As in \cref{fig:walking_spec}, the frequency components span over 1Hz to 4Hz and mask the HS, due to the jaw movement during speaking that creates vibrations and obscures the heart signals~\cite{bedri2015stick}.

% Next, we present our approach to overcome these challenges. \td{remove this sentene}

\section{Motion-resilient HR Estimation}\label{sec:method}
Typical signal processing techniques have shown effectiveness in HR estimation in the stationary case~\cite{InEarAudioWearableMeasurementOfHeartAnd}.
%: we adopt these approaches in this case also due to their low computational costs. 
However, they do not adequately isolate the HS from the corrupted audio under MAs. As previously discussed, motion-artefact elimination is a non-trivial problem. Typical signal processing techniques for denoising are more effective under certain signal-to-noise ratios (SNR) and errors increase with decreasing SNR~\cite{ali2015improved,ali2017denoising}. Additionally, the differences in the user's anatomy (different ear canal shapes, different earbud fit levels and thus changes in the resultant amplification) result in differences in the captured audio sounds, and this variability is poorly captured and processed using signal processing. 
Due to the recent successes witnessed by deep learning (DL) for denoising in numerous fields~\cite{lu2013speech,gondara2016medical}, we propose a novel pipeline using DL to eliminate MAs in audio signals and estimate HR.

% When capturing audio inside the ear canal, HS cannot be recorded during motion without simultaneously collecting MAs. This makes denoising of HS difficult as clean ground truth HS cannot be captured during motion.
% Similar to PPG-based denoising approaches~\cite{AccurateHeartRateMonitoringDuringPhysicalExercises}, we recorded ECG data simultaneously during the audio data collection, and use ECG as a reference signal providing a clean and high-quality heart signal under MAs. We developed neural networks to map the noisy HS signal to its corresponding ECG (a clean heart signal), and further estimated HR from the output \textit{synthesized ECG}. Our problem is thus framed as a denoising problem, but also as a synthesis problem. After the neural network is properly trained, we can directly employ it to obtain the clean heart signal for HR estimation. Therefore, ECG is only required for model training, but not for testing nor during real deployment of the technique.\td{remove this complete section? and add slightly in next section B when discussion ECG?}

In the following sections, we first present a signal processing approach for HR estimation, and then the proposed DL pipeline for MA removal.

\subsection{Signal Processing for HR Estimation}
The initial phase of our work involved the development of a signal processing pipeline for HR estimation. This aims to provide an efficient and computationally effective HR detection method, and to explore the potential of typical signal processing techniques in HR estimation under MAs. %\td{remove all the claims above as already mentioned at the beginning of this section. }%Different from~\cite{InEarAudioWearableMeasurementOfHeartAnd} which uses a peak detection technique on the pre-processed waveform, we detect HR from the frequency domain \td{because xxx}.

First, we compute the Hilbert transform of the audio to calculate the HR envelope. We then compute the spectrum of the envelope using Fast Fourier Transform (FFT) and detect the dominant peaks which are converted to the HR. This approach shows good performance on a clean and stationary signal (see \cref{sec:baseline}). However, when audio signals are corrupted with motions, dominant peaks in the spectrum correspond to motions, rather than the HR, thus introducing errors in HR estimation. More sophisticated denoising techniques are thus required to obtain clean HS under motion. The discrete wavelet transform (DWT) is therefore used to remove artefacts from the audio to isolate HS. Specifically, we filter out detail coefficients from the DWT based on signal variance, thus removing the noise components with a high variance from the mean.

Though denoising can yield a relatively clean HS signal, the denoised signals are still interfered by the MAs to some extent, due to the underlying complexity of the artefacts, and the closely overlapping frequency ranges of the artifacts and the HS. Therefore, we propose a frequency spectrum searching algorithm to estimate the HR from the denoised signal to account for the remaining MAs. Instead of searching the FFT peaks over the full frequency range of the denoised audio, we only search the HR peaks in a small frequency range corresponding to the range of allowable human HRs and the HR in the previous window. This guarantees that peaks in HR-unrelated frequency ranges are not taken as HR and the HR is temporally dependent on previous ones.

However, this system has limitations including error propagation due to temporal dependencies of the algorithm and a lack of robustness to changes in signal properties. It was also unable to reconstruct the clean audio, meaning that the data could not be used for metrics other than heart rate. Thus, we acknowledge that a more sophisticated approach to the problem, specifically to addressing signal denoising, is required.

\subsection{Overview of the Deep Learning-Based Pipeline}

% \begin{figure}[th]
%   \centering
%   \includegraphics[width=.8\linewidth]{figures/system_design/block.pdf}  
% \caption{hEARt system flowchart.\dm{This figure can be straightened to a thin fashion and arranged in two columns.}}
% \label{fig:system_flowchart}
% \end{figure}

\begin{figure*}[t]
  \centering
  \includegraphics[width=1\linewidth]{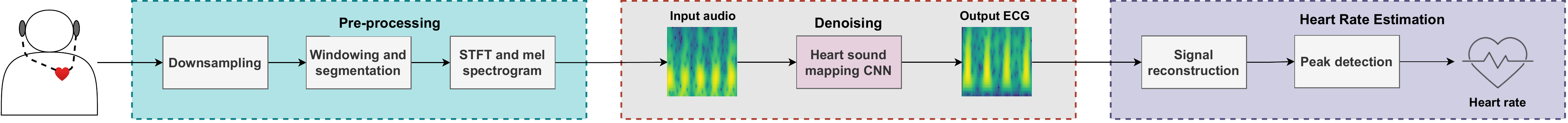}  
\caption{hEARt system flowchart.}
\label{fig:system_flowchart}
\end{figure*}

An overview of hEARt, our designed motion-resilient HR monitoring system, is given in \cref{fig:system_flowchart}. Audio signals captured inside the occluded ear canal are used for HR estimation, which is performed in three stages: pre-processing, MA elimination and HR estimation. Pre-processing aims at removing the frequency components unrelated to HS. For MA elimination, we proposed a CNN-based network to map spectrograms of the noisy HS signal to spectrograms of its corresponding ECG (a clean heart signal) during the training phase, thus producing an output \textit{synthesized ECG}. Our problem is thus framed as a denoising problem, but also as a synthesis problem. %We use GT ECG signals as a reference since they are clean and high-quality heart signals, even under MAs.
We adopted a U-Net encoder-decoder architecture for denoising since audio (and specifically HS) is commonly represented in image form as spectrograms~\cite{UNetConvolutionalNetworksForBiomedicalImageSegmentation,HeartSoundClassificationUsingDeepLearningTechniques,TowardsTheClassificationOfHeartSoundsBased}. Initially developed for biomedical image segmentation, U-Net shows great potential in image denoising and super resolution~\cite{UNetConvolutionalNetworksForBiomedicalImageSegmentation,HighResolutionUNetPreservingImageDetailsForCultivated}. 
It captures important features in audio spectrograms via an encoder, and reconstructs the corresponding clean heart signal via salient representations via a decoder. More importantly, the skip connections in U-Net allow the reuse of feature maps to enhance the learning of the original information, making it suitable for denoising. Evidence also shows that U-Net performs well with limited training data, which matches our case~\cite{RecurrentResidualConvolutionalNeuralNetworkBasedOn}. Finally, HR is estimated using peak detection on the clean signals.

\subsection{Pre-processing}
The HS captured by the in-ear microphone are low frequency signals with a bandwidth of less than 50~Hz. To prepare the audio signals for processing, we downsample the audio from 22~kHz to 1~kHz and segment the audio into 2s windows, each with a 1.5s overlap with the previous window. 2s windows were selected to ensure there will always be multiple heart beats (at least 2) within a window, enabling the system to learn inter-beat properties. Each window is bandpass filtered between 0.5~Hz and 50~Hz using a fourth order %Infinite Impulse Response (IIR) 
butterworth filter to remove the DC offset and high frequency signals. This attenuates the frequency components not of interest for HR calculation, including music and ambient noise. Additionally, due to  occlusion of the ear canal, the majority of external noise is suppressed and not captured by the internally facing microphone. However, as outlined in \cref{sec:challenges}, MAs and other interfering signals lie overlapping frequency ranges with HS, therefore requiring additional processing. 

We process the GT similarly. The ECG, sampled at 130~Hz, is bandpass filtered between 10 and 50~Hz and upsampled to 1~kHz. The highpass cutoff for the ECG was selected to be 10~Hz as this was empirically found to emphasise the peaks in the ECG (the QRS complex) while attenuating the P and T waves (as seen in \cref{fig:sound_vs_ecg}). Since we are only interested in capturing the beats and the inter-beat timing (for measuring HR, and in future, HRV), only the QRS complex is of interest.

\subsection{Motion Artefact Elimination}

\subsubsection{Spectrogram Generation}

The MA elimination subsystem takes as input the pre-processed audio signals, and outputs the cleaned heart signals. To do so, it uses the GT ECG signal to supervise the denoising of the heart signals.
We compute log-mel spectrograms for the windowed audio and ECG signals using short-time Fourier transform (STFT), with a window size of 256 samples and hop length of 32 samples. 1024 FFT bins are used with zero padding and a Hann window. These parameters were empirically selected. Thereafter, the log-mel spectrogram is computed using 64 mel bins. Log-mel spectrograms were chosen over spectrograms since they provide more detailed information in the low frequency region, where HS frequencies reside. The resulting log-mel spectrogram is a 64x64 matrix for each window. Since audio is captured in both ears, a spectrogram is computed for each channel and these are stacked together to form one 64x64x2 input. The output is a single channel ECG spectrogram. The spectrograms are normalised between 0 to 1, to aid network training. Normalisation is carried out for all spectrograms by dividing by a constant value, to maintain the difference in the signal amplitude for different activities.

\subsubsection{Network Structure}

\begin{figure}[ht]
    \centering
    \includegraphics[width=0.98\linewidth]{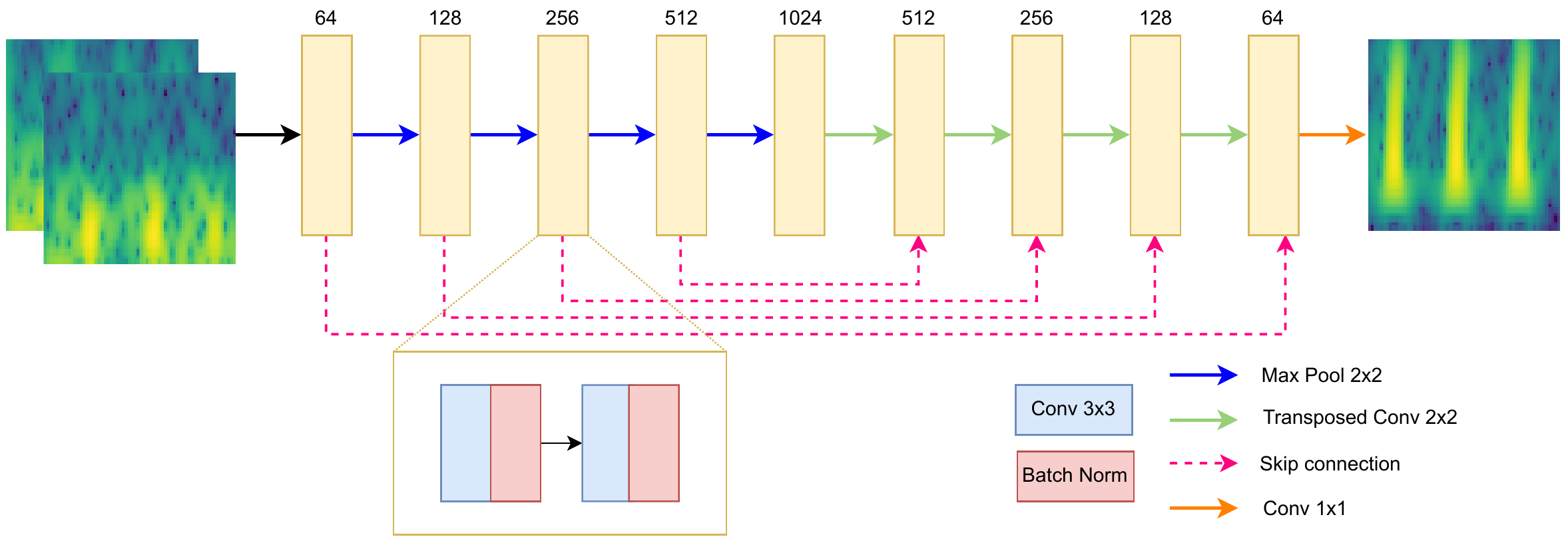}
    \caption{U-Net autoencoder architecture.}
    \label{fig:unet}
\end{figure}

\cref{fig:unet} provides the architecture of the U-Net used for denoising. In the encoder (or contraction path), the model consists of repeated 3x3 convolutions (with a ReLU activation function), batch normalisation and max pooling blocks with a stride of 2 to downsample the data. After pooling, dropout is applied with a rate of 0.1 to avoid overfitting. Each time the data is downsampled, the number of feature maps is doubled to enable the network to learn complex structures in the data. In the decoder (expansion path), the data undergoes successive up-convolutions where the number of feature maps is halved at each step. After each up-convolution, the feature maps are merged with the corresponding feature map from the encoder and then undergo the convolution and batch normalisation layers as in the encoder. In the final layer, a 1x1 convolution is used to map the final feature maps into a single 64x64 output image. 

\subsubsection{Transfer Learning}
On account of the small dataset, transfer learning is used to improve the results of the HS denoising. To achieve this, the model is pre-trained using the PASCAL HS dataset~\cite{pascal-chsc-2011}, where log-mel spectrograms of HS are used as both input and label to the network. By doing this, we aim to improve the ability of the network to extract representative audio features and encodings related to HS. The pre-trained model weights are set as the initialization weights for the CNN, which is fine-tuned using our data. This helps leverage and transfer the knowledge learnt about HS using PASCAL, as well as avoiding overfitting on a small dataset.

\subsubsection{Training}
% The input audio spectrograms and their corresponding ECG spectrograms are used to train the network. Due to the limited size of the dataset, five fold cross validation is done whereby the network is trained on 16 users and tested on 4 users, thus ensuring that no user appears both in the training and test set. The model is trained empirically for 100 epochs using the Adam optimiser with a learning rate of 0.01 and a batch size of 64.
% When choosing the training parameters, our objective was to strike a good balance between performance and computation complexity.

The input audio spectrograms and their corresponding ECG spectrograms are used to train the network. We use leave-one-out cross validation for testing whereby each subject is held out as the test-set and a model trained on the other 19 users. The model is trained empirically for 100 epochs using the Adam optimizer with a learning rate of 0.001 and batch size of 64.
When choosing training parameters, our objective was to strike a good balance between performance and computational complexity.

The system uses mean square error (MSE) or L2 loss (\cref{eqn:mse}). This loss minimises the distance between the GT ECG spectrogram ($y_{ij}$) and the noisy audio spectrogram ($\hat{y}_{ij}$), where $i$ and $j$ represent the time and frequency index respectively, and $T$ and $F$ represent the total number of bins over the time and frequency dimensions respectively.

\begin{equation}
MSE = \frac{1}{TF}\sum(y_{ij}-\hat{y}_{ij})^2
    \label{eqn:mse}
\end{equation}

\subsubsection{Signal Reconstruction}
We convert the reconstructed clean spectrograms to time-domain waveforms for HR estimation. The Griffin-Lim algorithm~\cite{FastGriffinLimAlgorithm} is used for spectrum inversion due to its ability to reconstruct signals from spectrograms without phase information. The converted waveforms are then merged into a continuous time-series signal by averaging the overlapping regions.

\subsection{Heart Rate Estimation}\label{sec:hr_est}
Heart rate estimation is performed in an 10s long window, where each window has a 5s overlap with the previous window~\cite{RobustHeartRateEstimationFromMultipleAsynchronous}. Each window undergoes the Hilbert transform to compute the envelope of the signal. Thereafter, a Gaussian moving average filter smooths out small ripples and peaks in the signal. Peak detection is calculated on the resultant signal, and the timings between consecutive peaks are used to compute the average heart rate for the window. Finally, a moving average window of 5 samples is used to remove outliers from the predictions.

\section{Implementation}
In this section we present the implementation details of our system, describing our prototype and the methodology we followed to run our data collection campaign.

\subsection{Prototyping}
\begin{figure}[ht]
%   \centering
  % include fourth image
  	\begin{subfigure}[t]{.69\columnwidth}
  	\centering
  	% include fourth image
  	\includegraphics[width=0.83\linewidth]{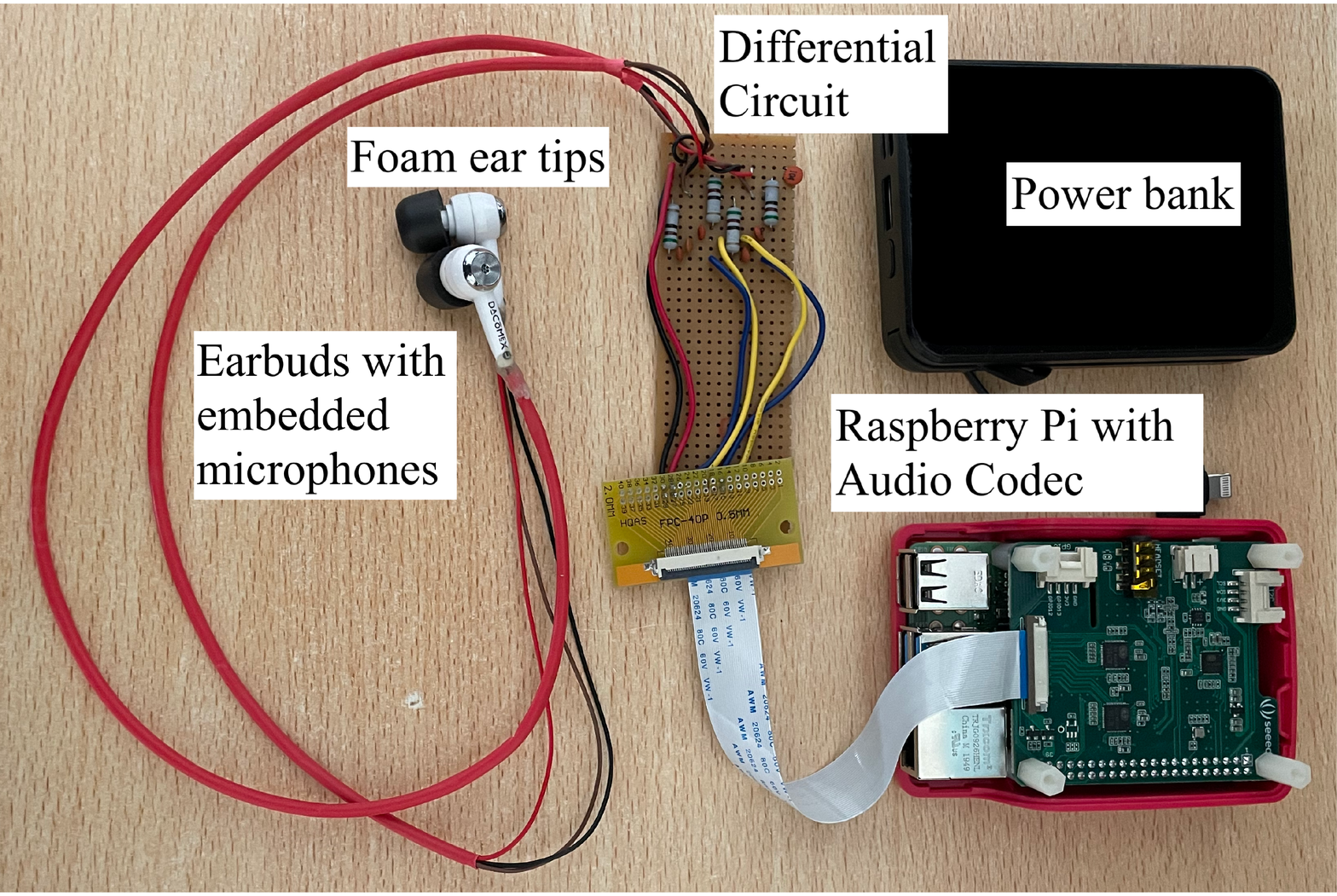}  
  	\caption{}
  	\label{fig:prototype_1_1}
  \end{subfigure}
  \begin{subfigure}[t]{.30\columnwidth}
  	\centering
  	% include first image
  	\includegraphics[width=0.97\linewidth]{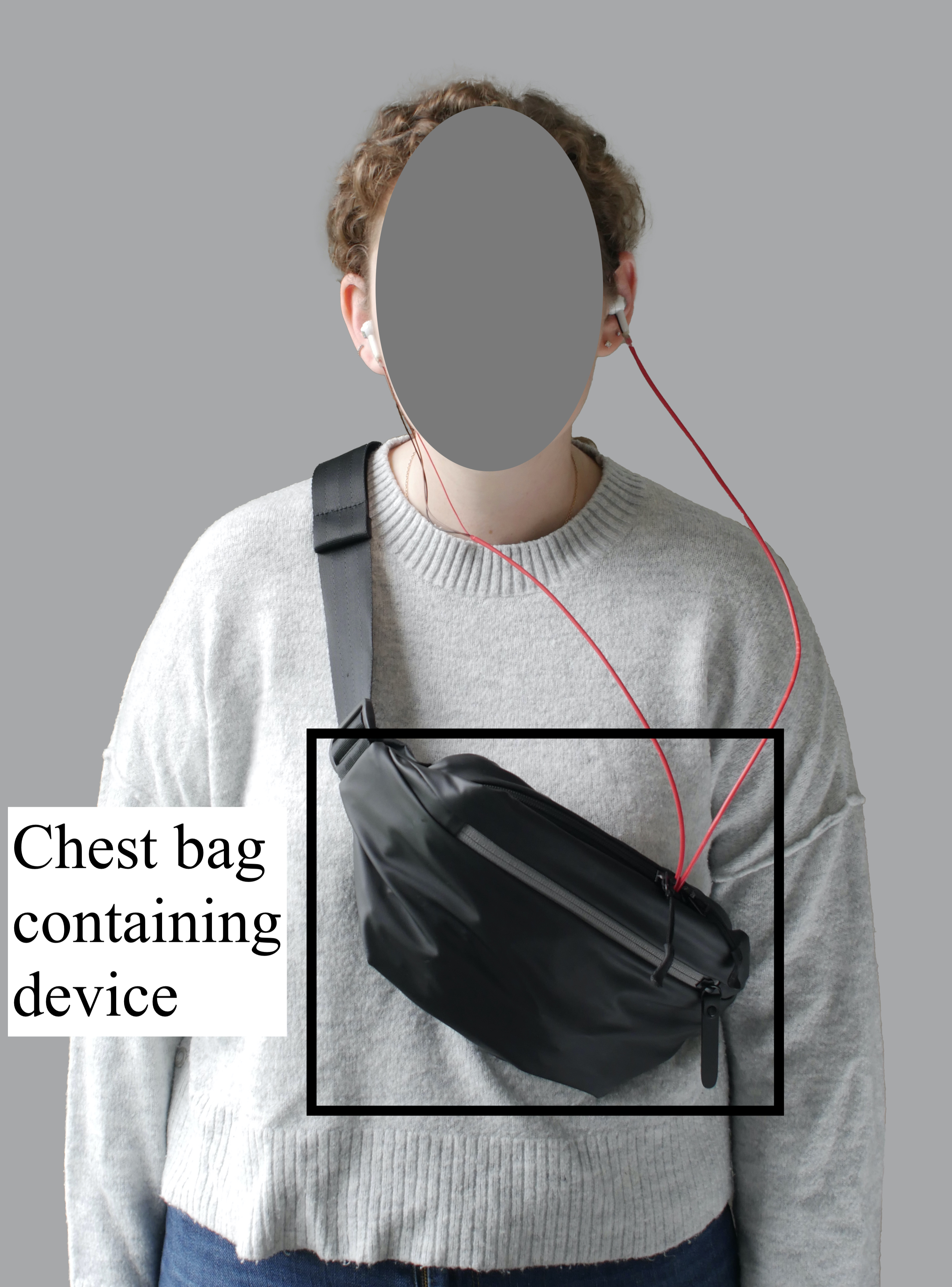} 
  	\caption{}
  	\label{fig:prototype_2_1}
  \end{subfigure} 
\caption{(a) Prototype and (b) participant wearing the device.}
\label{fig:prototype}
\end{figure}

Although in-ear microphones have been integrated into existing commercial earbuds (e.g., AirPods Pro), no API is available to access the microphone output. To gather data and understand the potential of our approach, we developed our earbud prototype by customizing existing earbuds, as shown in \cref{fig:prototype}. Specifically, we embedded two analogue omnidirectional MEMS microphones (SPU1410LR5H-QB from Knowles~\cite{datasheet:SPU1410LR5H-QB}) into a pair of wired earbuds, %(Dacomex Binaural Intra-Earphone), 
as shown in \cref{fig:prototype_1_1}. The microphones were selected due to their flat frequency response from 10~Hz to 10~kHz which encompasses the frequency range of HS, speech and MAs. The microphones were connected to a differential circuit for common mode rejection of power line noise and other noise sources and then sampled by an audio codec (ReSpeaker Voice Accessory Hat~\cite{ReSpeaker4MicLinearArrayKitForRaspberry}) onto a Raspberry Pi 4B. To make the system portable, the circuitry and Raspberry Pi were placed in a chest bag which was worn by the participants during the experiments (\cref{fig:prototype_2_1}). This ensured that the device did not interfere with the participant's natural movement while undergoing the tasks. 

\begin{figure}[ht]
  \centering
  % % include fourth image
  % 	\begin{subfigure}{.23\textwidth}
  % 	\centering
  % 	% include first image
  % 	\includegraphics[width=.99\linewidth]{figures/prototyping/rubber_unfilt_a.pdf}  
  % 	\caption{}
  % 	\label{fig:rubber_filt_short}
  % \end{subfigure} 
  %  \begin{subfigure}{.23\textwidth}
  % 	\centering
  % 	% include first image
  % 	\includegraphics[width=.99\linewidth]{figures/prototyping/foam_filt_a.pdf}  
  % 	\caption{}
  % 	\label{fig:foam_filt_short}
  % 	\end{subfigure}
    \begin{subfigure}{.24\textwidth}
  	\centering
  	% include first image
  	\includegraphics[width=1\linewidth]{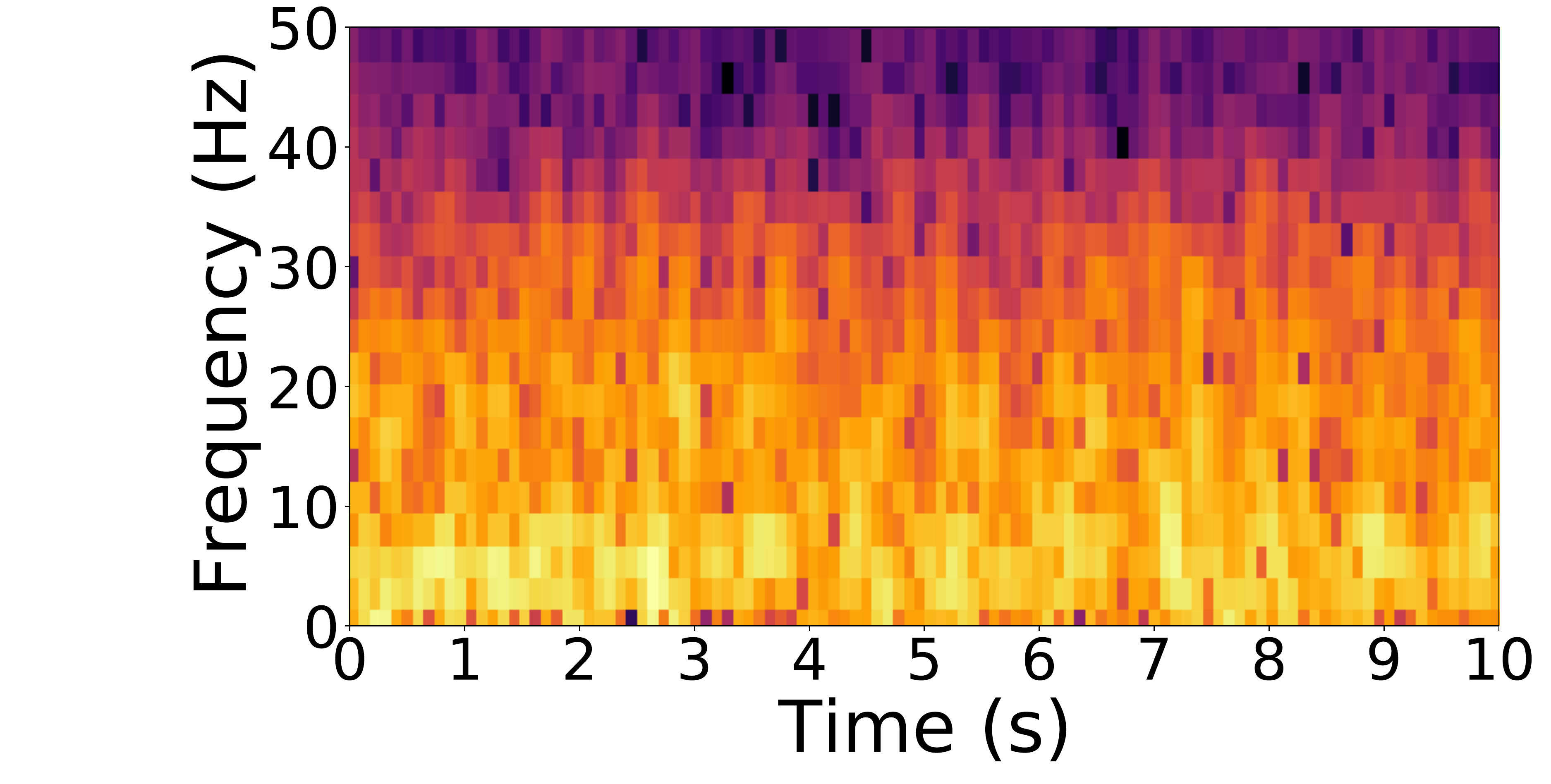}  
  	\caption{}
  	\label{fig:rubber_spec}
  \end{subfigure} 
   \begin{subfigure}{.24\textwidth}
  	\centering
  	% include first image
  	\includegraphics[width=1\linewidth]{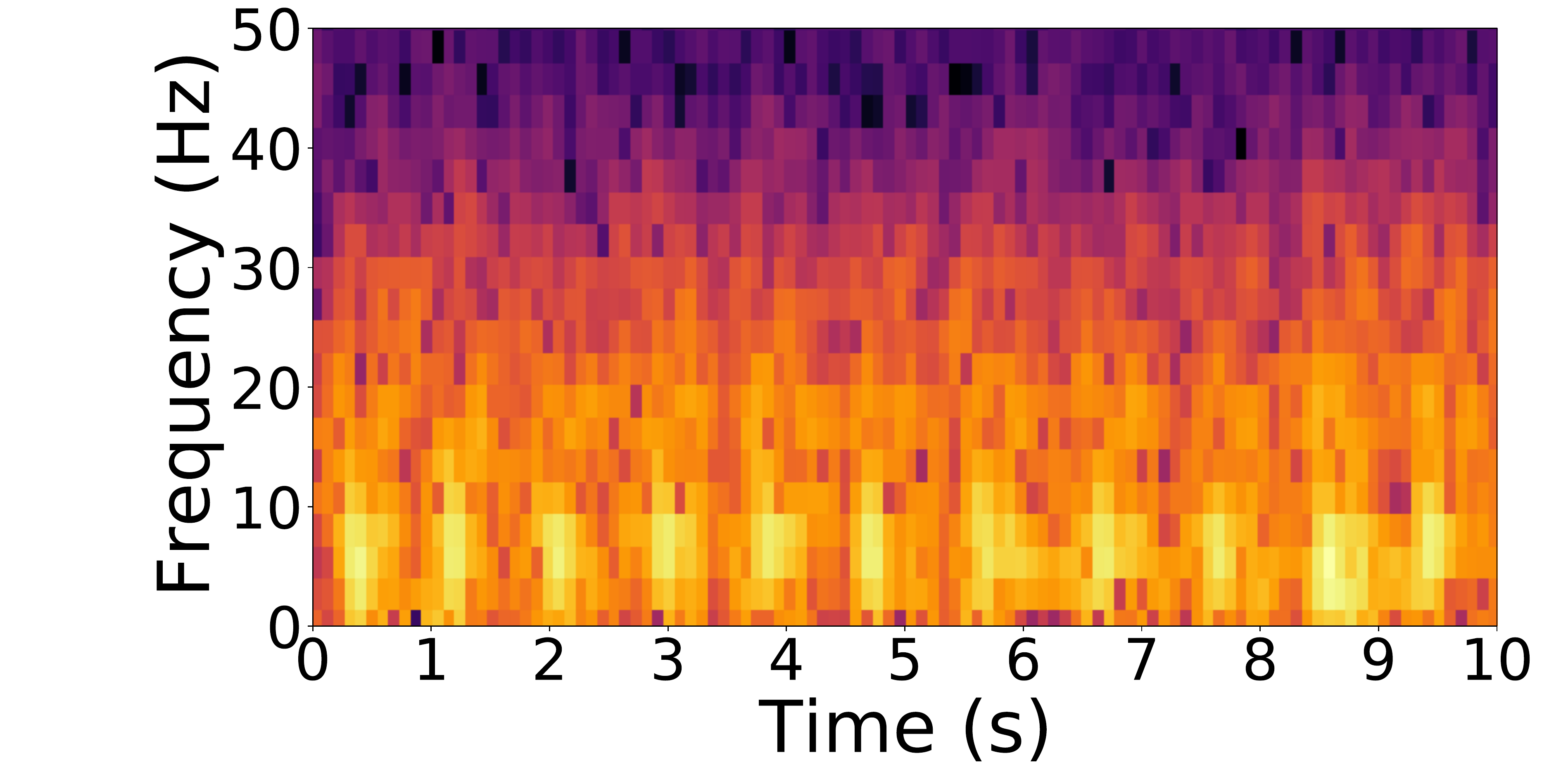}  
  	\caption{}
  	\label{fig:foam_spec}
  	\end{subfigure}
\caption{Comparison of signals collected when occluding the ear-canal with (a) a silicon ear tip and (b) a foam ear tip.}
\label{fig:occlusion_silicon}
\end{figure}

Although the occlusion effect implies the possibility of detecting bone-conducted sounds from the ear canal, measuring HS with an in-ear microphone is extremely challenging. Unlike walking, which generates strong vibrations, heart beat movement is subtle, resulting in very weak HS. As shown in \Cref{fig:rubber_spec}, when using the earbud equipped with a silicon ear tip, it is difficult to identify heart beats from the signal. We overcome this challenge by replacing the silicon ear tip with a foam ear tip, which (1) largely suppresses/absorbs external sounds, resulting in a lower noise floor; (2) ensures a better sealing of the ear canal, thereby winning more amplification gain from the occlusion effect (shown in \Cref{fig:foam_spec}). With this upgrade, our prototype is able to measure HS with good SNR.

\subsection{Data Collection} \label{sec:data_collection}
We use an ECG chest strap (Polar H10~\cite{PolarH10PolarGlobal}) to measure the GT heart beat signal. We extract the raw ECG from the Polar H10 and use it as both the clean heart signal for the CNN and to calculate the GT HR. The microphone data was sampled at 22050~Hz and the ECG at 130~Hz. Due to the difference in the sampling rates, there is a maximum of a 150ms delay between the audio and the ECG signal. However, since HR estimation is performed in 10s windows, this delay is negligible. We synchronized the data by aligning the timestamps of the ECG signal with the timestamps of the audio file.

% \begin{figure}
%   \centering
%   % include fourth image
%   \includegraphics[width=.8\linewidth]{example-image-a}  
% \caption{Participant wearing the device}
% \label{fig:wearing_prototype}
% \end{figure}

We invited 20 participants (13 males and 7 females) for data collection\footnote{The experiment has been approved by the Ethics Committee of the institution.}. In addition to the stationary case, we considered three activities that are regarded as active or, that, because of their nature, interfere with the in-ear microphone: walking, running, and speaking. These activities were also selected as they match the conditions in~\cite{ferlini2021ear} used to study in-ear PPG.
While wearing our earbud prototype and the ECG chest strap, the participants first kept stationary for 30~seconds to obtain the reference HR. Then, they performed each of the tasks continuously for 2 minutes~\cite{MeasuringHeartRateDuringExerciseFromArtery}. The 30~second stationary reference is only used for the baseline signal processing approach, and is not used in the hEARt system. 
When performing the walking and running activities, participants were allowed to pick a comfortable pace, and were instructed to move freely within a 5x4~m area. 
For the speaking activity, they were given a passage to read out loud. 
We processed a total of 160 minutes of in-ear audio corresponding to four tasks across all participants. Data collection was done in the atrium of a busy building, and as such data was collected in the presence of uncontrolled ambient noise, including human speech, the opening and closing of doors and low frequency power grid hum and air conditioning. 
The data collected while running for participants 2 and 14 was excluded on account of poor data quality. This occurred because one of the earbuds fell out during the intensive running activity. As such, there was no seal between the ear and the earbud meaning that the occlusion effect could not be leveraged.

The distribution of GT HR varies per activity. While stationary, the mean HR is 70$\pm$12~BPM, with a minimum and maximum of 45 and 114~BPM. While walking, the HR ranges from 51 to 129~BPM with a mean HR of 86$\pm$14~BPM. Running has the highest average HR (109$\pm$23~BPM) with the largest range of HR (50 to 187~BPM). The HR while speaking is similar to that while stationary with a mean of 76$\pm$12~BPM and a range of 51 to 124~BPM. 

% The distribution of the GT HRs for each activity are provided in \cref{tab:hr_breakdown}. From the table it is evident that the smallest variation in HR is found in the stationary cases, with the variation increasing as activity level increases. Likewise, the mean HR is highest for running and lowest when stationary.
% \cm{table can be cut and described more in text?}
% \begin{table}[ht]
%     \centering
%     \caption{Heart rates recorded with the Polar H10 for the four activities.}
%     \resizebox{0.8\columnwidth}{!}{%
%     \begin{tabular}{|c||cccc|}
%     \hline
%     \textbf{Activity} & \textbf{Mean} & \thead{\textbf{Standard}\\ \textbf{Deviation}} & \textbf{Minimum} & \textbf{Maximum} \\
%     \hline
%     \hline
%     % Full dataset & 93 & 25 & 48 & 171 \\
%     Stationary & 70 & 12 & 45 &114\\
%     Walking & 86 & 14 & 51 & 129\\
%     Running & 109 & 23 & 50 & 167\\
%     Speaking & 76 & 12 & 51 & 124\\
%     \hline 
%     \end{tabular}
%     }
%     \label{tab:hr_breakdown}
% \end{table}

\section{Performance Evaluation}

\subsection{Metrics}
We evaluated the performance of our system according to the following metrics~\cite{Ismail2021HeartReview}:

    % MAE
    \textbf{(i)} Mean Absolute Error (\textbf{MAE}): the average absolute error between the GT HR ($BPM_{true}$) and the calculated HR ($BPM_{calc}$) for each window ($i,i \in [1,N]$). %(\cref{eqn:mae}).
    % \begin{equation}
    %     MAE = \frac{1}{N}\sum _{i=1}^{N}|BPM_{calc}(i)-BPM_{true}(i)|
    % \label{eqn:mae}
    % \end{equation}
    %  MAPE
    
    \textbf{(ii)} Mean Average Percentage Error (\textbf{MAPE}): the average percentage error over each window. % (\cref{eqn:mape}).
    % \begin{equation}
    %     MAPE = \frac{1}{N}\sum _{i=1}^{N}\frac{|BPM_{calc}(i)-BPM_{true}(i)|}{BPM_{true}(i)}*100
    % \label{eqn:mape}
    % \end{equation}
    % % SAE
    % \item \textbf{SAE}: the standard deviation of MAE.
    % % SAPE
    % \item \textbf{SAPE}: the standard deviation of MAPE.
    % BA
    
    \textbf{(iii)} \textbf{Modified Bland-Altman plots}: a scatter plot indicating the difference between the two measurements (i.e. the \textit{bias} or error) for every true value (i.e. HR from the GT). A modified Bland-Altman (BA) plot is constructed so that 95\% of the data points lie within $\pm$1.96 standard deviations of the mean difference between the methods~\cite{Giavarina2015UnderstandingAnalysis}. 
    BA plots are used clinically to assess the level of agreement between two measurement methods~\cite{Giavarina2015UnderstandingAnalysis}. In this work, we compare the calculated HR to the GT HR for each 10s window. %\td{is equations 2 and 3 ok to be removed and only give reference to save space?}
% \end{itemize}

\subsection{Baseline Comparison}\label{sec:baseline}

\Cref{tab:baseline_mape_ppg} shows the performance comparison between the proposed DL based hEARt system and two signal processing approaches - (1) the proposed signal processing (SP) method (referred to as SP) leverages the DWT for signal denoising and extracts HR from the frequency spectrum of the denoised signals. 
(2) we additionally compare our methods to the baseline developed by Martin and Voix~\cite{InEarAudioWearableMeasurementOfHeartAnd} (referred to as baseline), which uses Hilbert transforms and peak detection for HR estimation in the time domain, under stationary conditions.
Our proposed SP approach outperforms the baseline significantly for stationary and running, and marginally for walking and talking. This demonstrates that the baseline algorithm designed for stationary is unable to generalize to motion conditions, and an additional denoising module is required.
% \dm{you justify baseline for specific for stationary, but its performance still worse than SP. Can we say the results demonstrate two things (1) HR detection in time domain leads to lower performance even in stationary, (2) baseline is not generalizable to motion scenarios. } 
Comparing the SP with hEARt, we observe that hEARt outperforms SP for each of the activities, showing that the DL based technique is better at generalizing to the differences in the data than the SP approach. While performance in the stationary case is comparable, with more intense motion interfering with the HS, SP fails to capture the HR from the signal and the performance severely deteriorates. hEARt outperforms SP significantly with a relative improvement of 51\%, 54\% and 48\% for walking, running and talking respectively, suggesting the effectiveness of hEARt in HR estimation. Additionally, errors for the stationary, walking and running conditions are less than 10\%, meaning that the system is accurate by ANSI standards for these activities~\cite{PhysicalActivityMonitoringForHeartRateANSI}.

The results for speaking are noticeably the worst of the four activities studied. This is consistent with \cref{fig:talking_spec}, where it is clear that speaking brings more severe noises than the other activities. Perhaps against intuition, this is not on account of speech being detected by the microphone since the frequencies of \textit{audible} human speech are significantly higher than those of interest in the hEARt system. Rather, speaking causes movement of the jaw and head, and deformation of the ear canal due to jaw movement. These movements result in low-frequency bone-conducted vibrations which could be interpreted as heart beats. They are also non-periodic and random in nature and are thus harder to remove, resulting in higher errors. This is in contrast to walking and running which are largely periodic and more homogeneous and thus easier to remove.

\cref{tab:baseline_mape_ppg} also compares the performance of hEARt with that of in-ear PPG (as studied by Ferlini et al~\cite{ferlini2021ear}). It is evident from the table that (i) although PPG is the gold-standard for HR measurement, full-body motion causes significant degradation in HR measurement quality and (ii) our audio-based approach performs better than in-ear PPG. We thus believe that in-ear audio could be used as an alternative to, or in combination with, in-ear PPG for HR measurement through the ear. 

\begin{table}[ht]
    \centering
    \caption{Comparison between hEARt, the two baselines and in-ear PPG in terms of MAPE (\%).}
    \resizebox{0.95\columnwidth}{!}{%
     \begin{tabular}{|c||ccc|c|}
     \hline
     \textbf{Activity} & \thead{\textbf{hEARt}} & \thead{\textbf{Signal} \\ \textbf{Processing}} & \textbf{Baseline~\cite{InEarAudioWearableMeasurementOfHeartAnd}}& \thead{\textbf{In-ear}  \\ \textbf{PPG~\cite{ferlini2021ear}}}\\
     \hline
     \hline
     Stationary & \textbf{4.32 $\pm$ 3.99}& 4.93 $\pm$ 8.33 &9.88 $\pm$ 6.93& ---\\
     Walking & \textbf{9.53$ \pm$ 8.28}& 19.41 $\pm$ 16.03 &20.90 $\pm$ 11.22& 27.14\\
     Running & \textbf{9.80 $\pm$ 7.93} & 21.43 $\pm$ 15.30 & 34.28 $\pm$ 8.73& 29.84\\
     Speaking & \textbf{12.06 $\pm$ 8.88} & 23.37 $\pm$ 9.39  &24.23 $\pm$ 7.98& 12.52\\
     \hline
     \end{tabular}
     }
    \label{tab:baseline_mape_ppg}
\end{table}

\subsection{hEARt Overall Performance}

\begin{figure}[ht]
\centering
\begin{subfigure}{.24\columnwidth}
  \centering
  % include 1 image
  \includegraphics[width=1\linewidth]{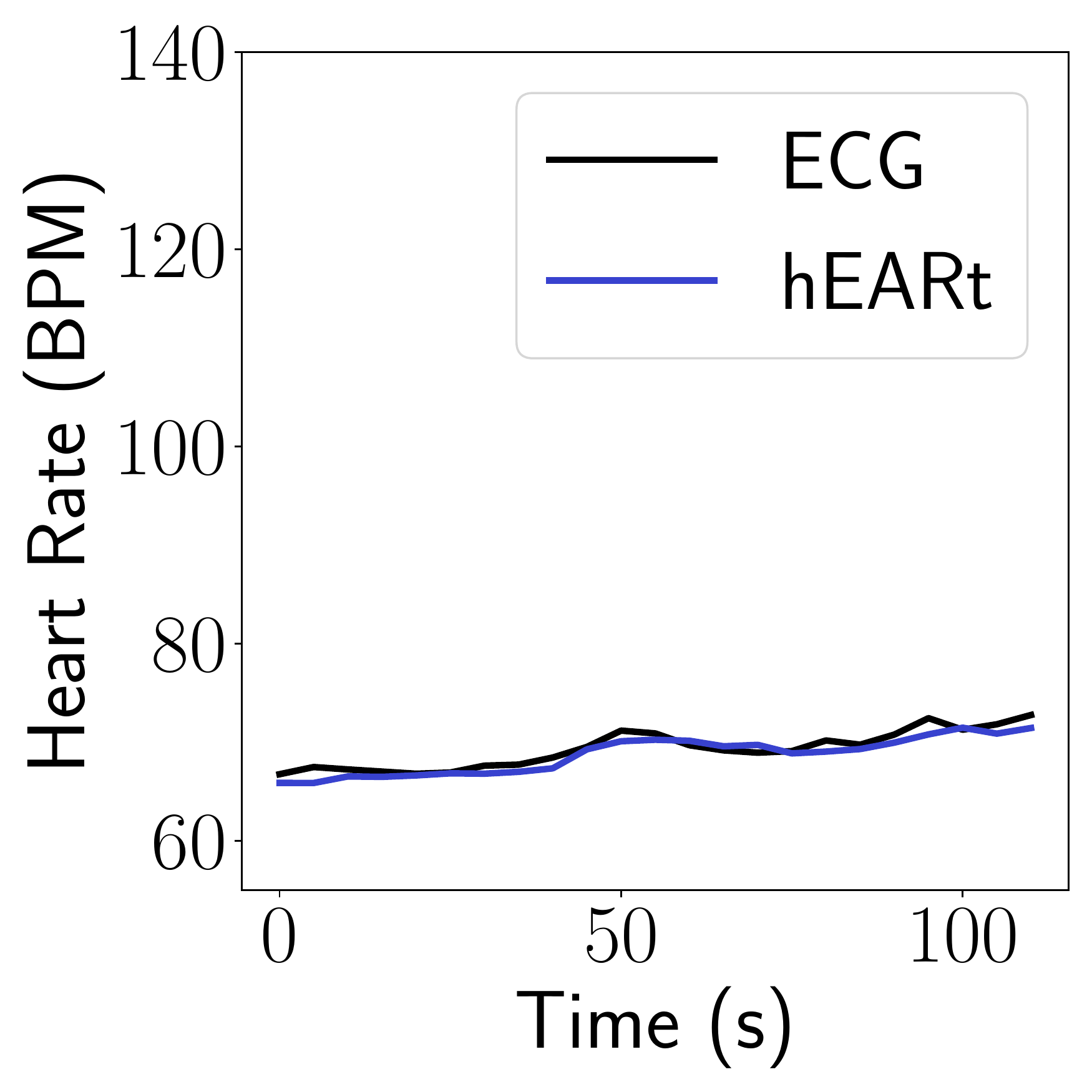}  
  \caption{Stationary}
  \label{fig:hr-sitting}
\end{subfigure}
\begin{subfigure}{.24\columnwidth}
  \centering
  % include 2 image
  \includegraphics[width=\linewidth]{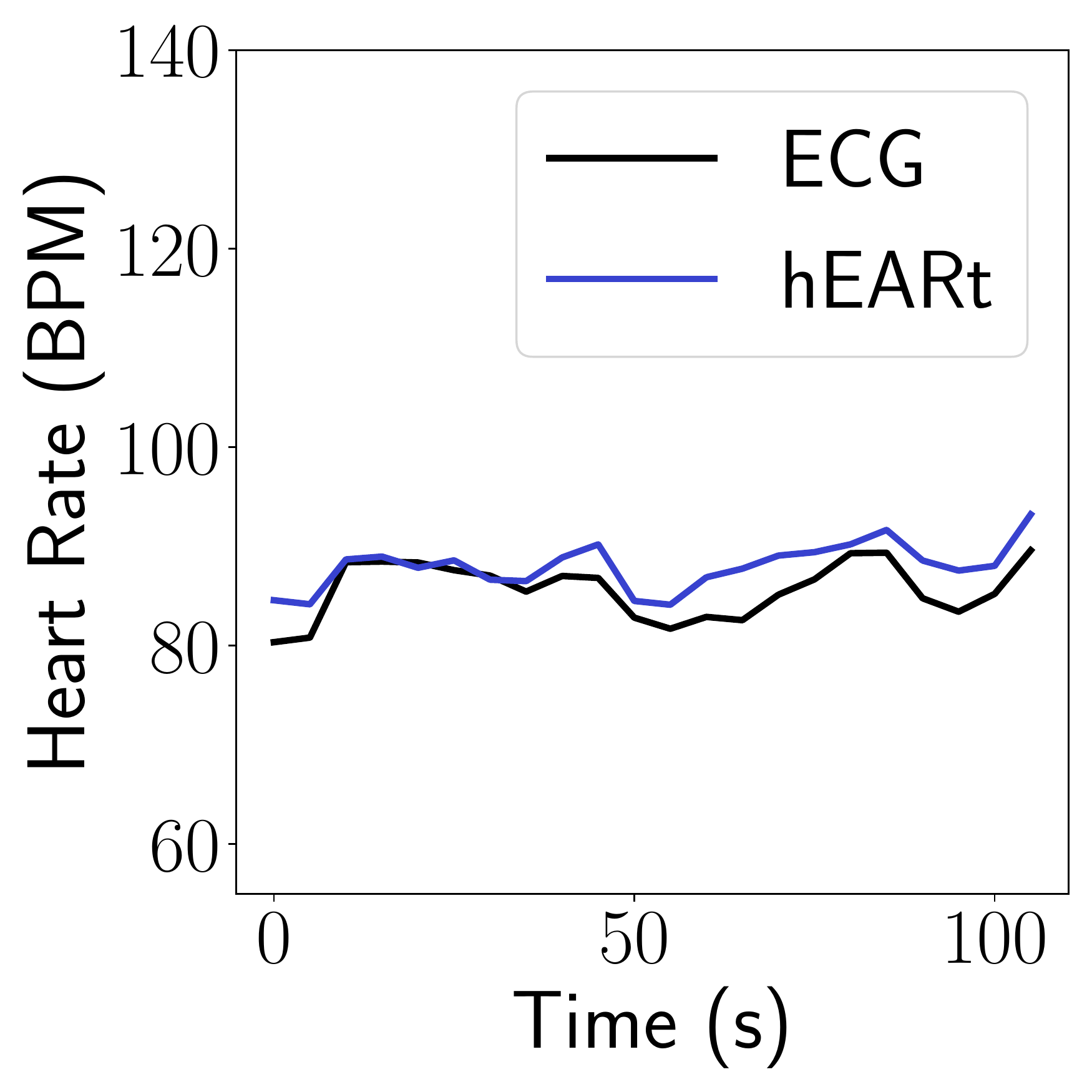}  
  \caption{Walking}
  \label{fig:hr-walking}
\end{subfigure}
\begin{subfigure}{.24\columnwidth}
  \centering
  % include second image
  \includegraphics[width=\linewidth]{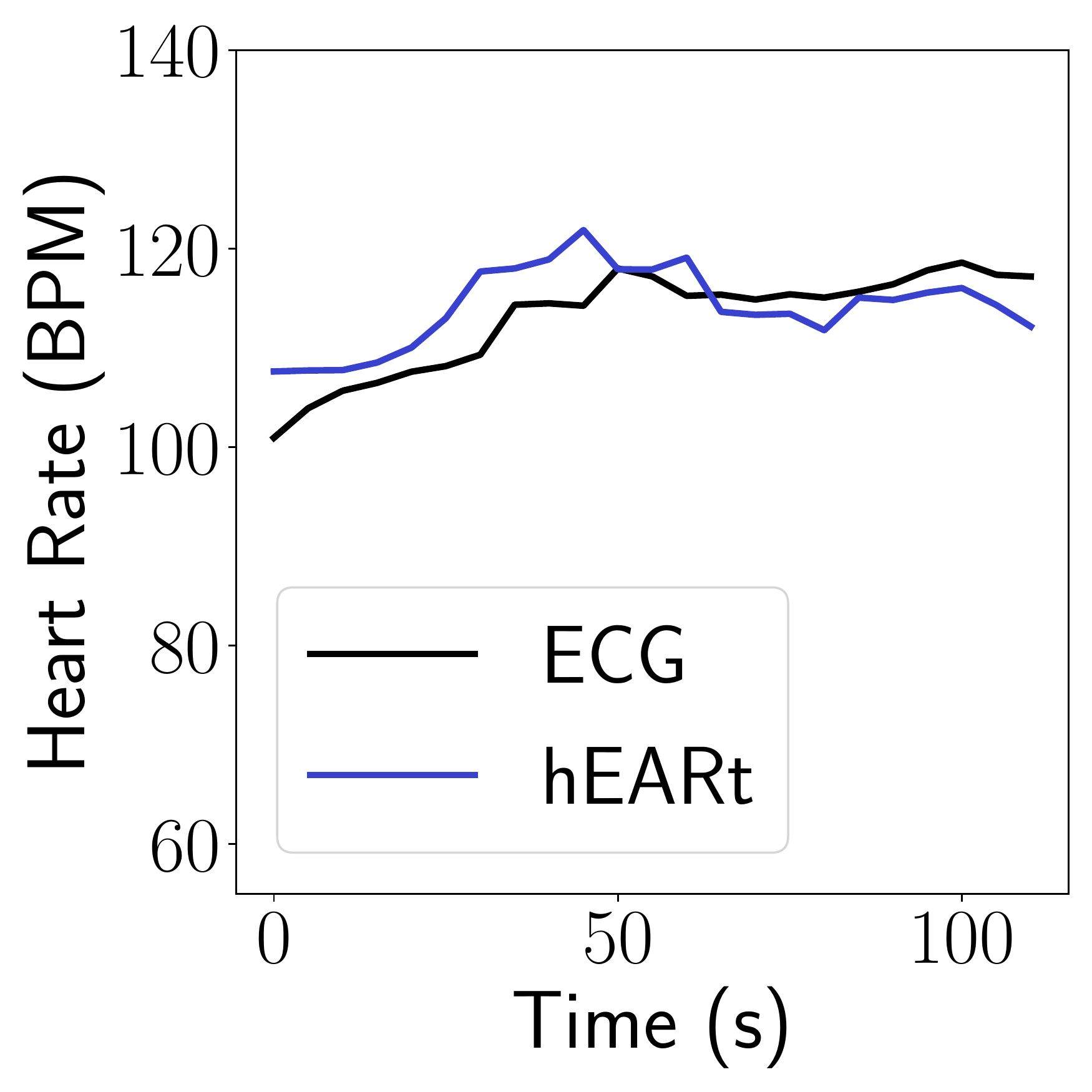}  
  \caption{Running}
  \label{fig:hr-running}
\end{subfigure}
\begin{subfigure}{0.24\columnwidth}
  \centering
  % include second image
  \includegraphics[width=\linewidth]{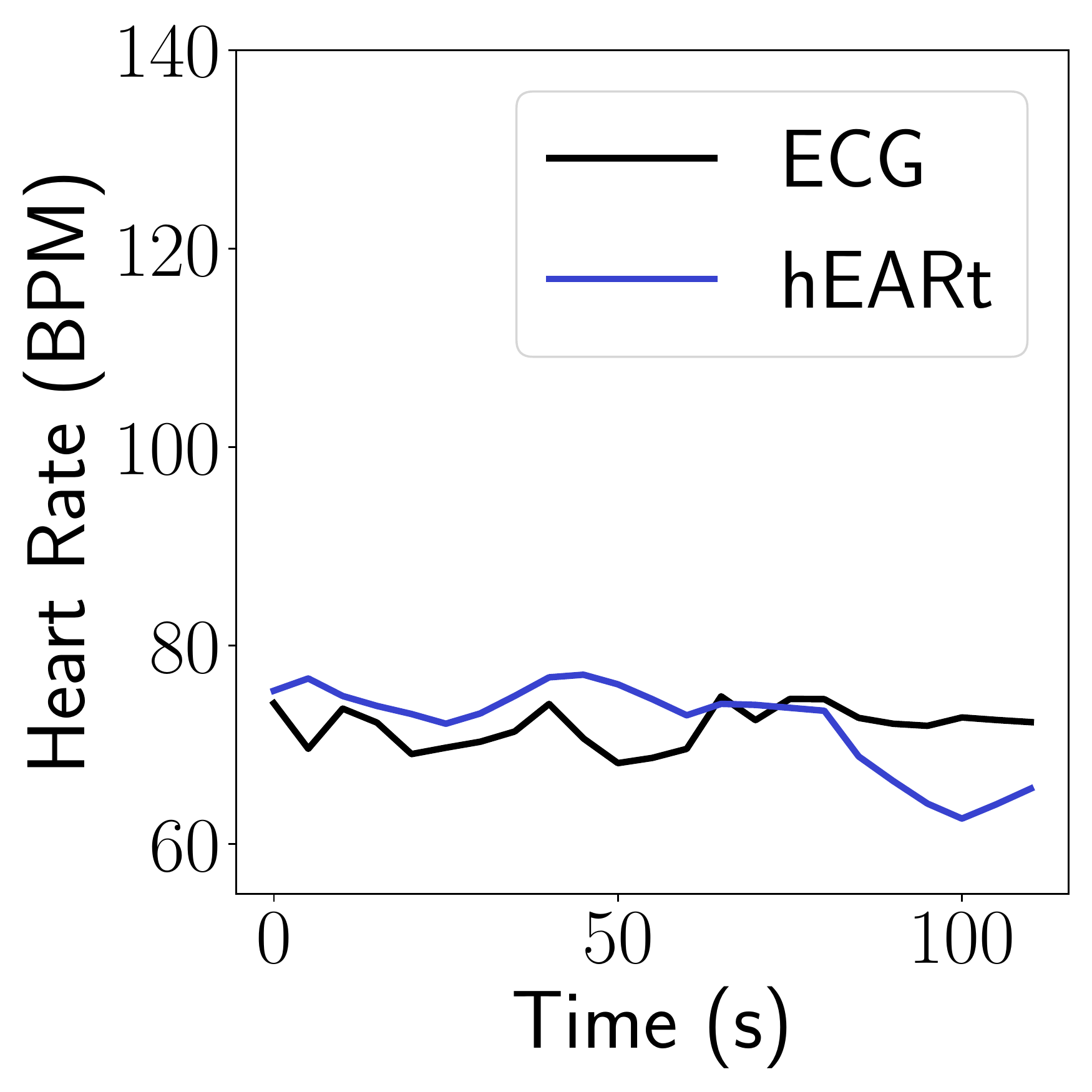}  
  \caption{Speaking}
  \label{fig:hr-talking}
\end{subfigure}
% \vspace{-0.5\baselineskip}
\caption{Qualitative longitudinal performance of heart rate extraction under different activities.}
\label{fig:hr_vs_time}
\end{figure}

\cref{fig:hr_vs_time} shows the qualitative assessment of hEARt in tracking HR over time. 
We compared the GT HR collected via ECG chest-strap with the one extracted from the in-ear audio for one participant over the four different activities. It can be observed that the proposed approach is able to accurately and continuously track the user's HR during the four activities (stationary, walking, running, and speaking), suggesting the potential of in-ear audio for HR estimation under MA.
For speaking, the larger error is due to jaw movements. However, the overall trend of estimated HR still aligns with GT.

Overall, the system achieves a MAE of 3.02 $\pm$ 2.97~BPM, 8.12 $\pm$ 6.74~BPM, 11.23 $\pm$ 9.20~BPM and 9.39 $\pm$ 6.97~BPM for stationary, walking, running and speaking respectively. As noted, we achieve the lowest performance during speaking in terms of MAPE as shown in Table 2.
Concretely, given an average heart rate of 76~BPM (the mean HR while talking as per \cref{sec:data_collection}), a MAPE of 12.06\% means our system misses (or adds) about 0.15 heart beats every second, or, 1 heart-beat every 7 seconds. Similarly, the performance achieved for running is even more convincing: at an average heart rate of 109~BPM, we miscompute 0.18 heart beats per second, again amounting to around 1 heart-beat every 5 seconds.

\vspace{-0.3em}

\subsection{Individual HR Estimation}
\begin{figure}[ht]
    \centering
\begin{subfigure}[t]{0.46\columnwidth}
\centering
	\includegraphics[width=0.98\columnwidth]{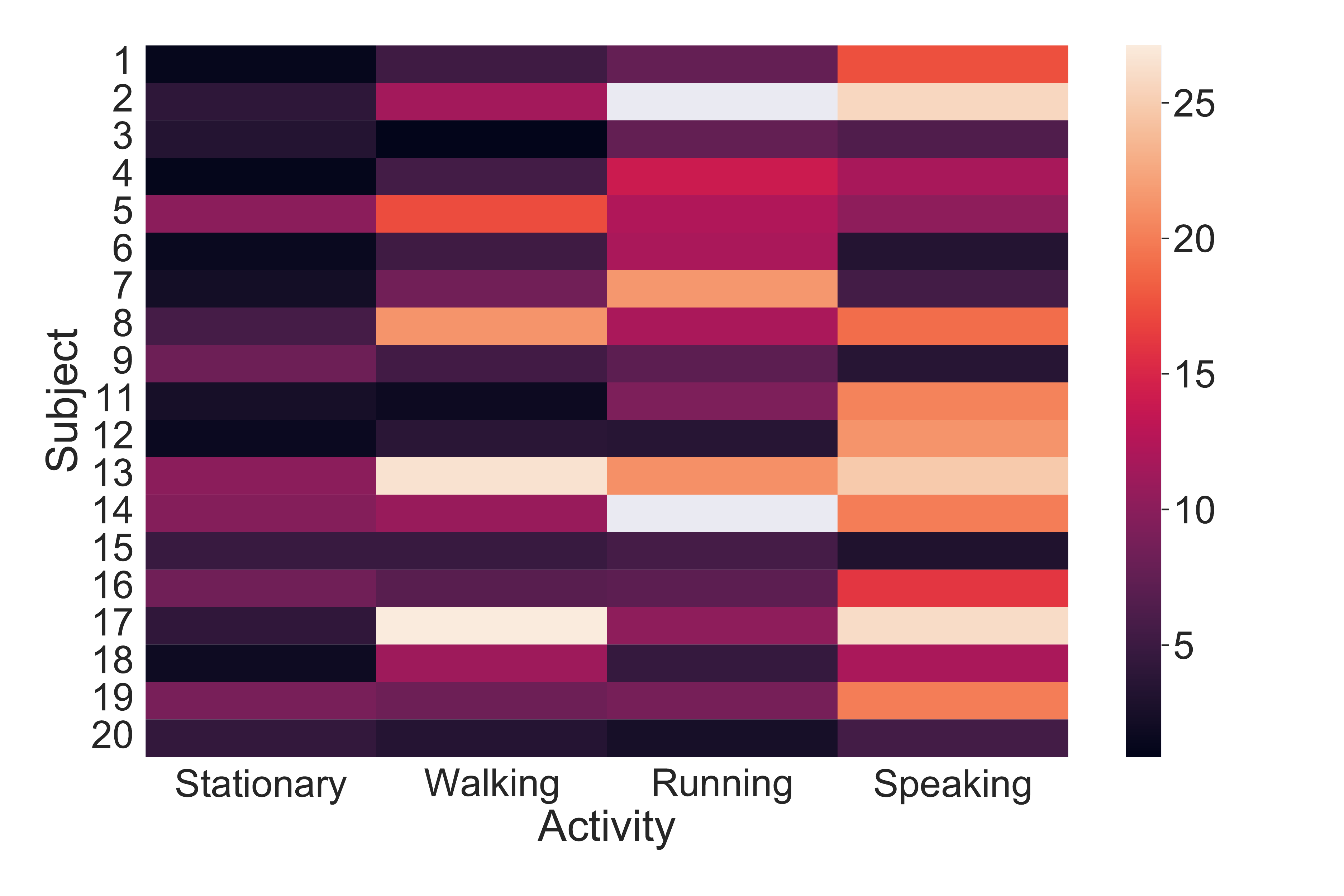}
       \caption{}
   \label{fig:heatmap}
\end{subfigure}
% \end{figure}
% \hfill
\centering
% \begin{figure}[h]
\begin{subfigure}[t]{0.46\columnwidth}
\centering
    \includegraphics[width=0.92\columnwidth]{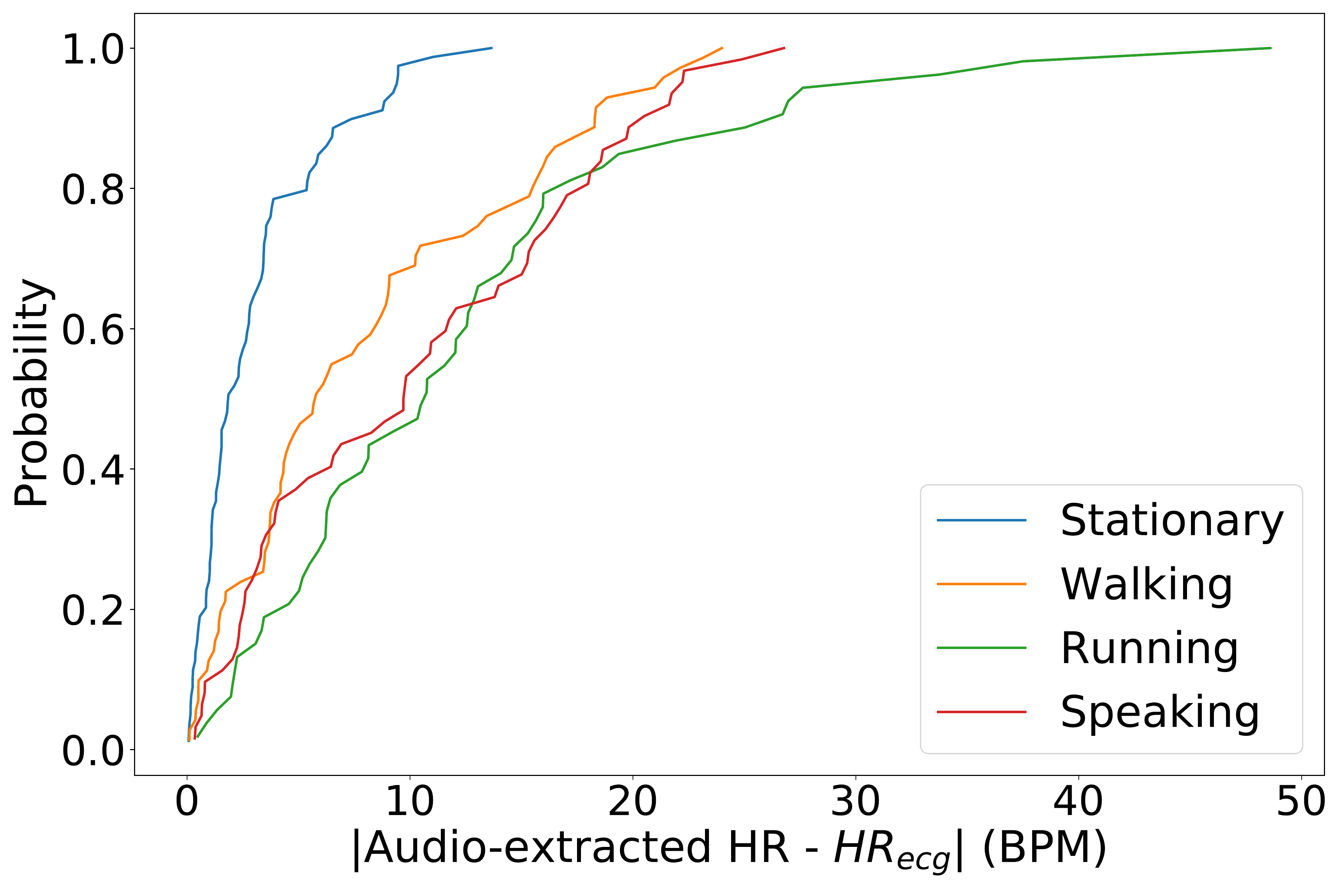}
    \caption{}
 	\label{fig:ecdf}
\end{subfigure}
% \end{figure}
% \hfill
% \vspace{-0.11in}
\caption{(a) MAPE heatmap per subject and (b) empirical CDF.}
% \label{fig:ecdf}
% \caption{Population statistics over the four activities.}
\label{fig:statistics}

\end{figure}

Next, we evaluate our approach under different activities for all subjects. 
First, we provide some insights on the population statistics.
\cref{fig:heatmap} reports a heatmap of the MAPE of the audio-extracted HR for every user across the activities.
Lighter colors correspond to greater MAPE values. Running for user 2 and 14 was removed due to a poor seal, and is represented as a white box (or NaN error). From the figure, we can extract a number of insights: \textbf{(i)} errors for motion conditions are higher than stationary. \textbf{(ii)} our system generalizes well to the different activities. \textbf{(iii)} One user experiences overall poor performance (user 13). This is due to a poorly fitting earbud, and poor quality GT data. \textbf{(iv)} Certain users experience poor performance in a specific activity (e.g. users 13 and 17 for walking). This is again likely due to an incorrectly fitting earbud in one ear which loosened during the activity, reducing the occlusion effect. These issues would be solved by the use of wireless earbuds (ensuring that the wires do not dislodge the earbuds during activity) and by ensuring a higher quality earbud fit. Overall, these results prove that the system is able to generalize to different users and that with high quality data, good HR estimation can be achieved.

To further understand the extent to which the various activities impact hEARt, for each of them we report the empirical cumulative distribution function (ECDF) of the error (\Cref{fig:ecdf}).
Looking at the ECDFs we can confirm what was observed in the heatmap. Specifically, our approach achieves an error of less than 12~BPM for over 60\% of users for all activities. As seen in the heatmap, most of the error observed comes from a few specific users rather than from the population in general. This performance on our academic prototype confirms that in-ear audio sensing of HR offers a promising alternative for continuous HR sensing in presence of motion. 

% As expected, when the intensity of the motion increases, we observe greater errors. 

% 13~BPM for over 60\% of the users when the user is walking or speaking (low intensity motion artefacts). The errors ramp up to about 20~BPM for more than 60\% of the study population whenever running is concerned.
% As expected, both walking and running present a similar trend, where our approach achieves an error of less than 15~BPM for over 60\% of the users. Notably, confirming our previous observations, this increases for speaking where the error for more than 60\% of the users is 20~BPM. 

% stationary - 60\% of the users, error is less than 10bpm. 

% walking - curve is less steep, shows that the error is increasing. 

\subsection{Bland-Altman Plots}\label{sec:ba_plot}

\begin{figure}[ht]
\centering
\begin{subfigure}{.24\columnwidth}
  \centering
  % include 1 image
  \includegraphics[width=1\linewidth]{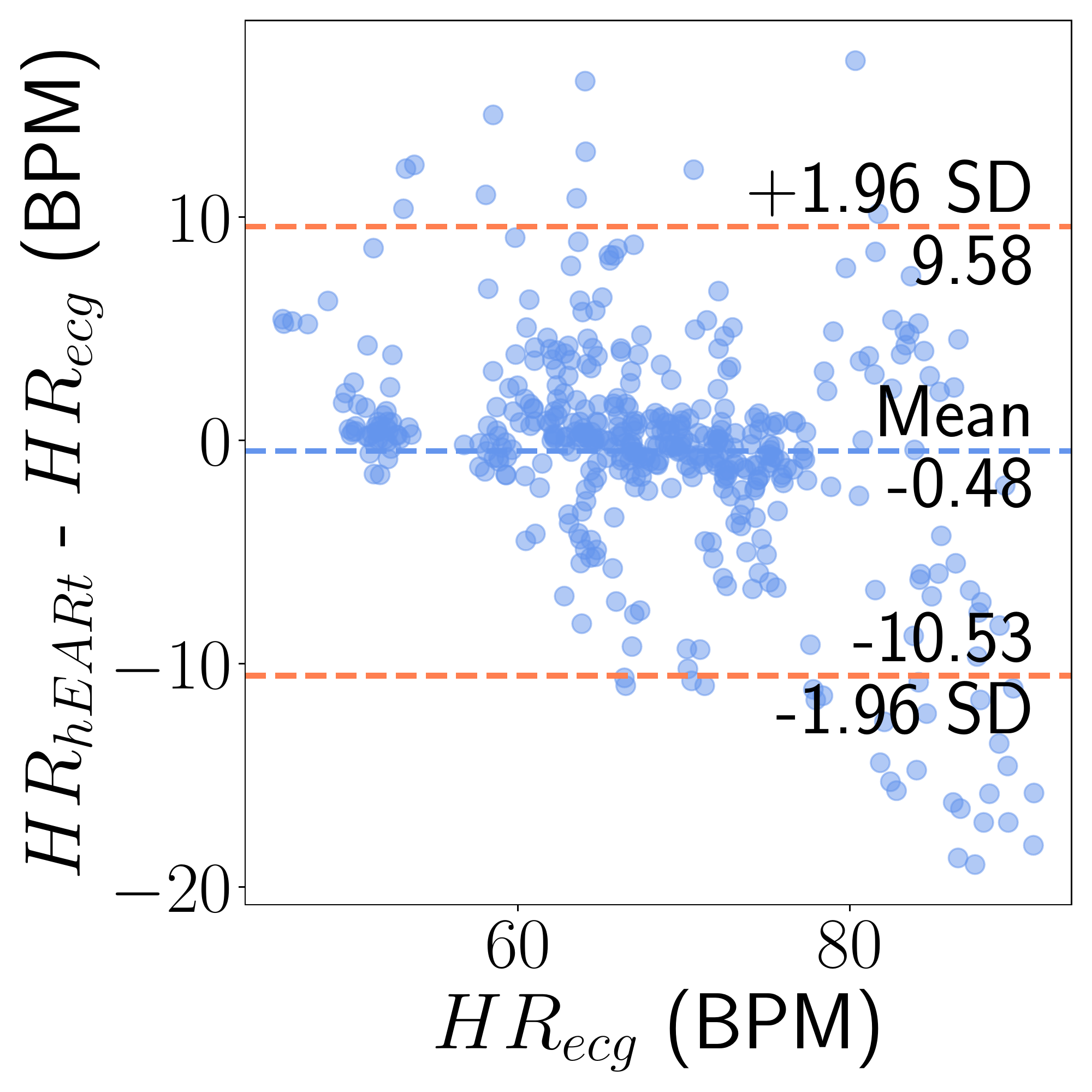}  
  \caption{Stationary}
  \label{fig:bland-sitting}
\end{subfigure}
\begin{subfigure}{.24\columnwidth}
  \centering
  % include 2 image
  \includegraphics[width=1\linewidth]{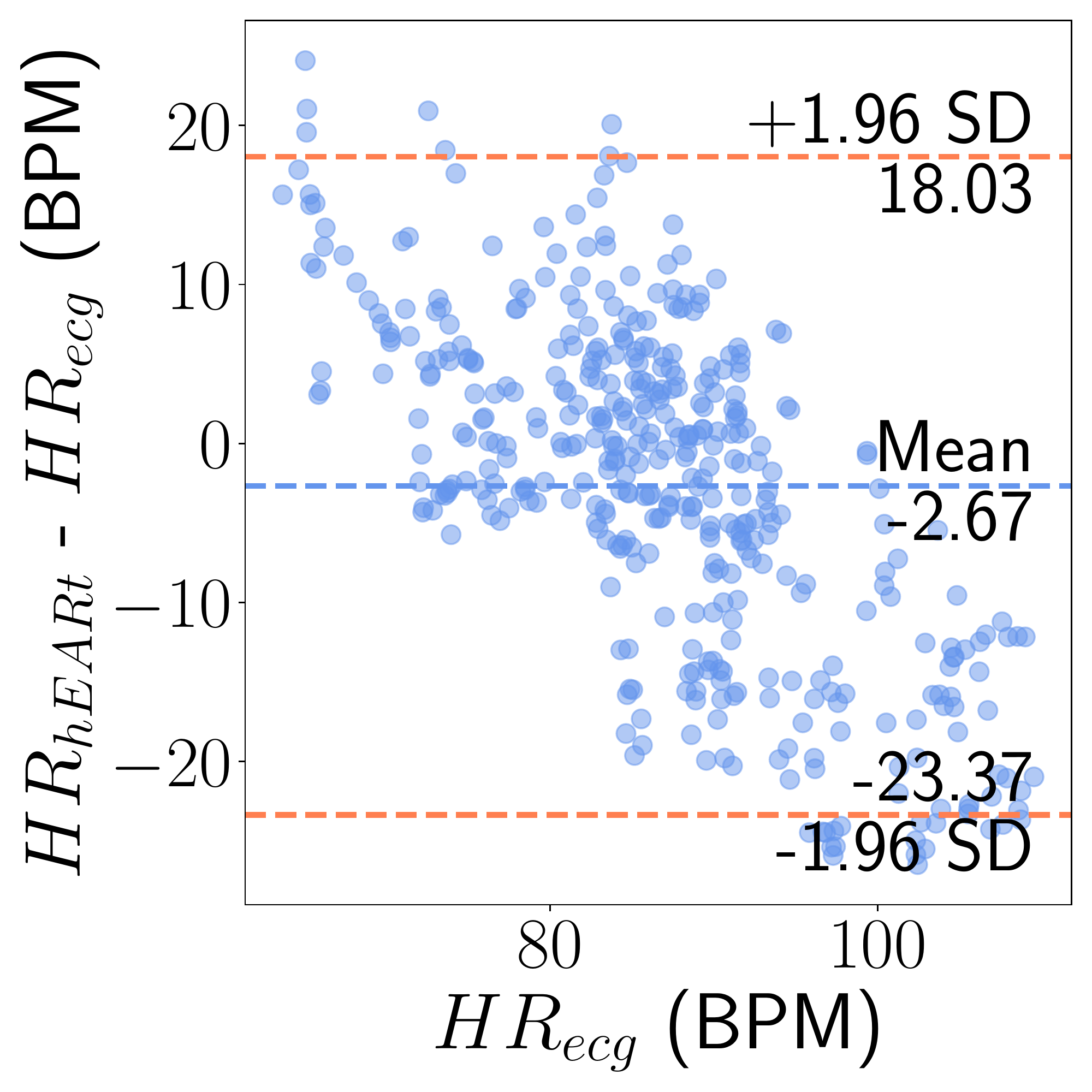}  
  \caption{Walking}
  \label{fig:bland-walking}
\end{subfigure}
\begin{subfigure}{.24\columnwidth}
  \centering
  % include second image
  \includegraphics[width=1\linewidth]{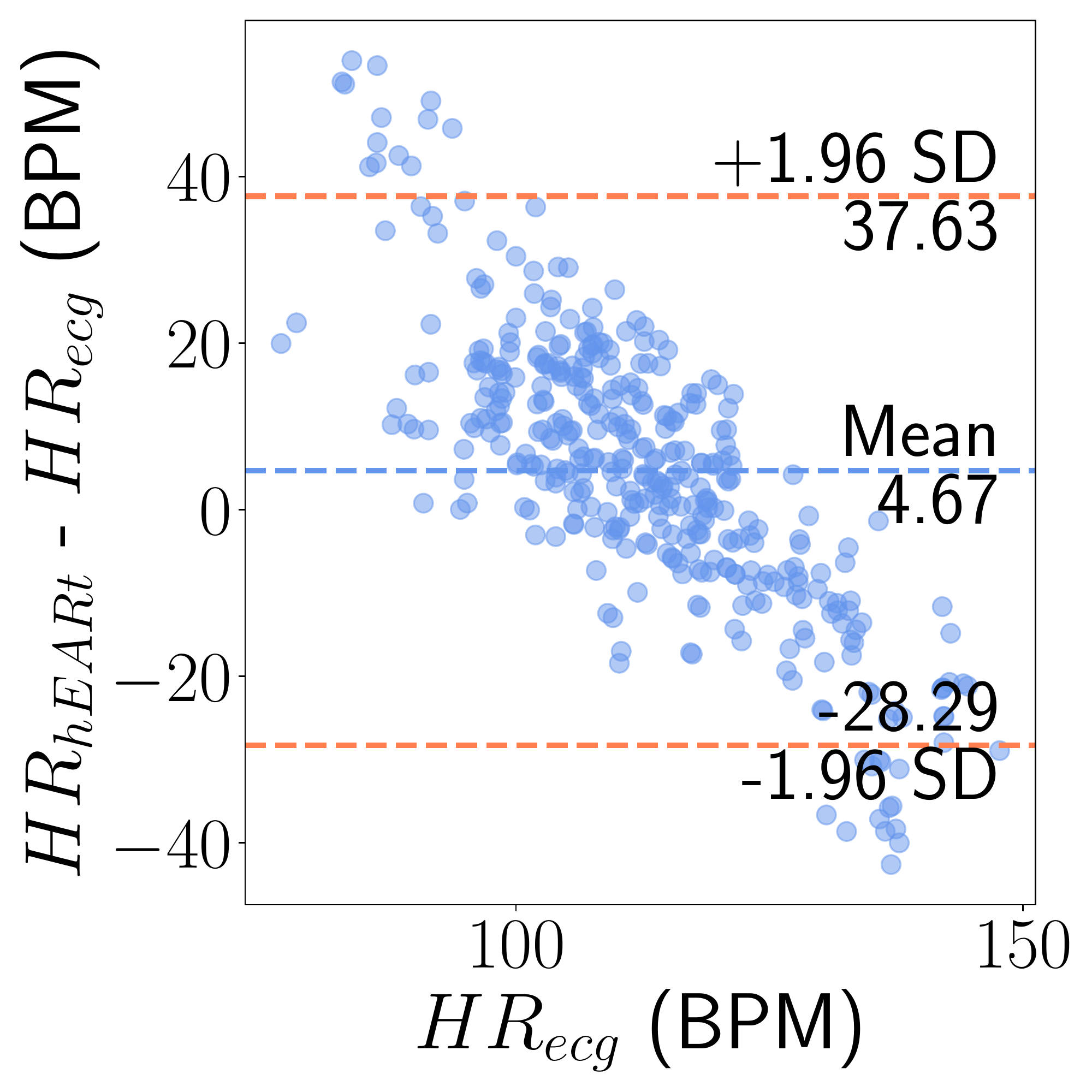}  
  \caption{Running}
  \label{fig:bland-running}
\end{subfigure}
\begin{subfigure}{0.24\columnwidth}
  \centering
  % include second image
  \includegraphics[width=1\linewidth]{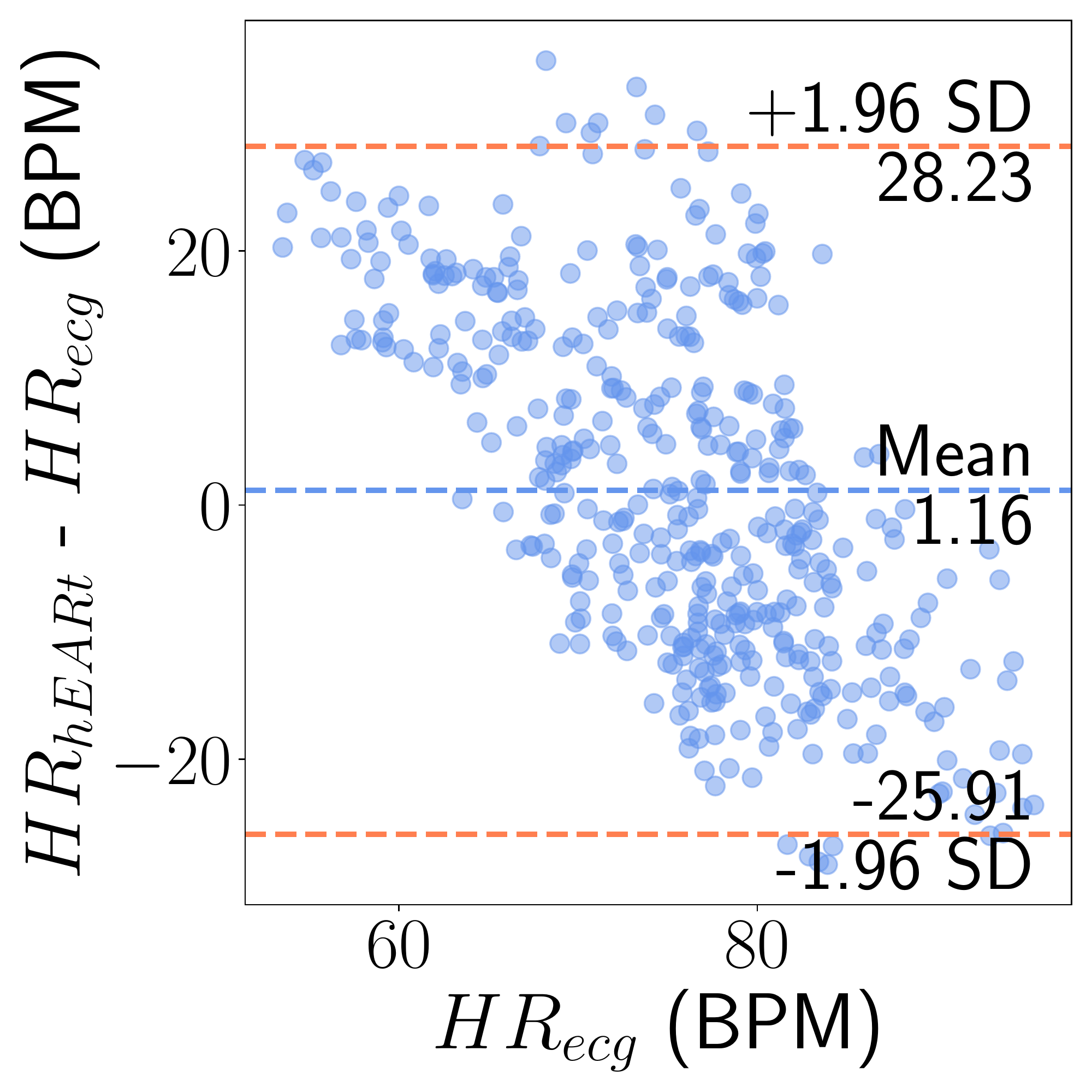}  
  \caption{Speaking}
  \label{fig:bland-talking}
\end{subfigure}
% \vspace{-0.5\baselineskip}
\caption{Modified Bland-Altman plot of heart rate extraction.}
\label{fig:bland-compare}
\end{figure}

To further analyze the results, we leverage modified BA plots.
% BA plots depict the difference between the HR computed from the in-ear audio and the HR reported from the ground truth on the y-axis ($Calculated - ECG_{HR}$) and the true value on the x-axis ($ECG_{HR}$). BA plots are commonly used to study the agreement between two measurements, allowing us to investigate how these differ (i.e., y-axis) for different values of HR (i.e., x-axis).
We report the BA plots (i.e. the agreement between the HR calculated with hEARt and that obtained from the GT chest strap) for each condition in~\Cref{fig:bland-compare}.
Specifically, \Cref{fig:bland-sitting} reports the agreement while stationary.
It is clear that the bias between the two measurements is minimal, with very low mean (only -0.48~BPM) and narrow limits of agreement (dashed red lines). Notably, the majority of the data points fall inside the limits of agreement, denoting the two measurements are in agreement.
On the other hand, with more intense activities like walking and running (\Cref{fig:bland-walking} and \Cref{fig:bland-running} respectively), 
wider limits of agreement are present, representing a greater standard deviation in the HR estimation. 
Interestingly, while overall the mean errors remain low (-2.67~BPM for walking and -2.41~BPM for running), our approach exhibits a larger error for estimation as HR increases. We observed this phenomenon both for walking (\Cref{fig:bland-walking}) and running  (\Cref{fig:bland-running}) motions. Notably, especially in the running case, this is observed when the frequency of the running overlaps with the HR values.
The spurious MA-induced spikes trigger a harsher response by hEARt that tries to remove the noisy peaks, thus leading to an underestimation of HR above 120~BPM. Additionally, another factor to explain the higher errors biased towards higher heart rates could be traced back to the imbalance of our dataset, where lower HR values are predominant.
Finally, in~\Cref{fig:bland-talking}, the mean error is again low but with fairly wide limits of agreement. This wide standard deviation again points towards the complexity of the speaking activity, meaning that extrapolating useful heart signals to compute HR from in-ear audio signals is a very challenging task, never tackled before.
Nonetheless, our approach still performs well.

\subsection{Long-term tracking performance}
\begin{figure}[ht]
    \centering
    \includegraphics[width=0.9\columnwidth]{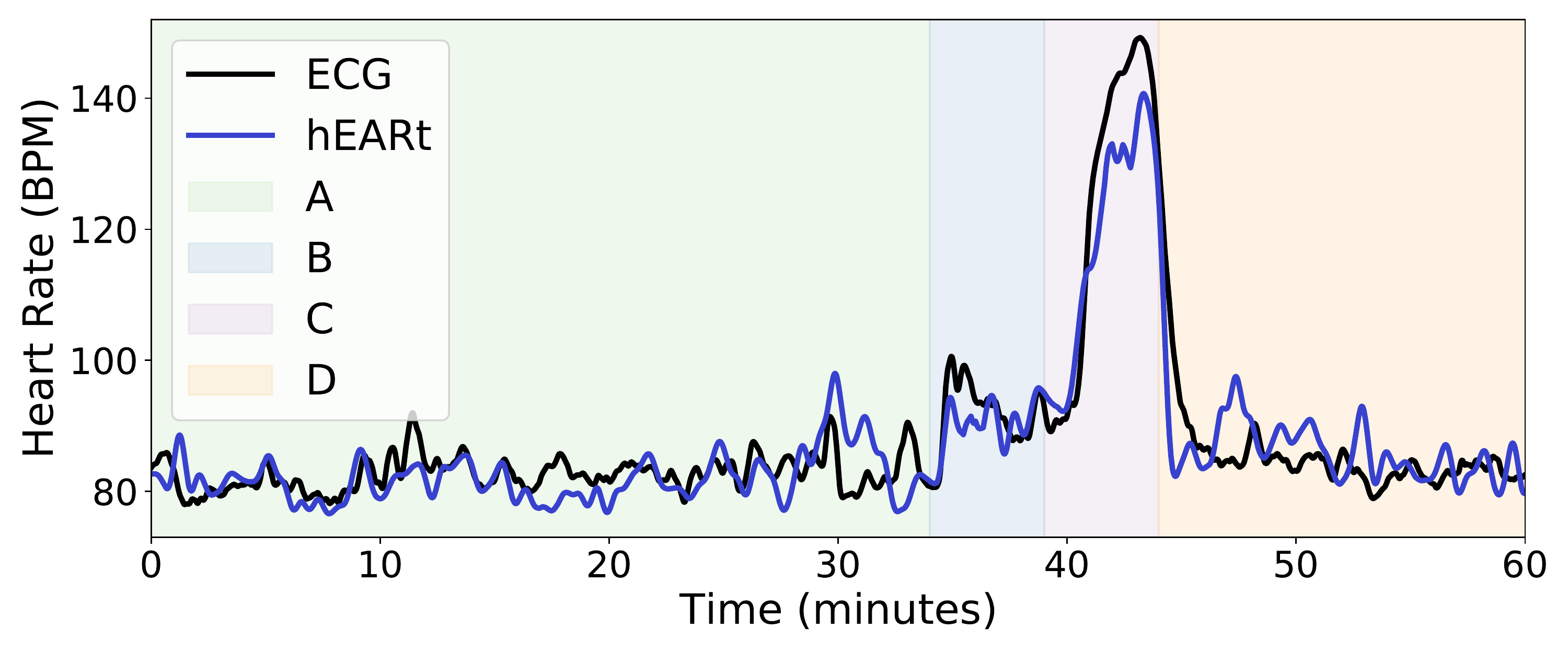}
    \caption{Longitudinal heart rate tracking. Coloured boxes indicate the different activities. A: Working while sitting. B: Walking. C: Running. D: Working while standing.}
    \label{fig:longitudinal}
\end{figure}

The results of the previous sections were obtained from experiments run under controlled conditions. To assess the real world effectiveness of the designed system, we collected an hour of data from one subject under conditions of daily life. During this time, the subject was instructed to undergo their activity as normal. This activity included working in an office, walking around, speaking (while working) and taking a short jog. The results of HR prediction for this study are provided in \cref{fig:longitudinal}. 
From the figure, it is evident that the system is able to accurately predict HR even in uncontrolled environments as the trends of the two lines match each other closely. However, as was seen in the BA plots in \cref{sec:ba_plot}, the system underestimates the higher heart rates. This is likely due to the distributions of heart rates in the dataset where the average HR is 85~BPM. %The average HR in the dataset is thus much lower than the maximum HR seen in the longitudinal study. This indicates that more training data at higher activities is needed for the network to generalize to all possible HRs. The underestimation could also be due to the filtering of the MAs being too aggressive in the presence of higher frequency signals as these normally represent noise.  
%\kb{I'm not sure if we should say this or not. They might then come back with a "why didn't you collect more data".}\dm{is underestimation in higher HR really true, given there is just a small period? In addition, if we follow this way, most of the low HR is overestimated, but we cannot use distribution to explain this. I am just wondering if underestimation high HR is a general problem and not related to our model, e.g., other paper also reports similar results. If so, we can just cite it. }\kb{I'm not sure about this. Maybe @Andrea has a better idea?}
%\cm{it would be good if Dong was right. i do not know what to suggest otherwise other than perhaps stop after "85BPM".}

The MAE of this longitudinal study is 4.96~BPM, which is a MAPE of 5.34\%. To further break this down, the MAPE of activities A, B, C and D are 4.41\%, 5.59\%, 9.51\%, and 5.94\% respectively. If we compare these results to those in \cref{tab:baseline_mape_ppg}, we can see that all activities have comparable performance to that of the controlled experiments. The results of this study prove that the model is generalizable to different conditions and to different activities. It also shows that the model is able to make accurate predictions even under conditions of mixtures of activities. Thus this study acts as a proof of concept of the in-the-wild feasibility of the hEARt system.

% \begin{figure}[ht]
%     \centering
%     \includegraphics[width=.7\columnwidth]{figures/results/sub_1_long_colours.pdf}
%     \caption{In-the-wild heart rate tracking. The coloured boxes indicate the different activities occurring during the study. A: Working while sitting. B: Walking. C: Running. D: Working while standing.}
%     \label{fig:longitudinal}
% \end{figure}

\subsection{Power and Latency Measurements}
To provide a full system analysis, we assess the power consumption and latency of the system implemented on a Raspberry Pi4. The trained hEARt CNN was converted to TensorFlow lite and deployed on the device. This mimics a stand-alone earable system whereby processing is done on device. \cref{tab:latency} provides a breakdown of the operation times for the various system components. Signal denoising was performed on a 2s window and HR extraction on an 10s window, as detailed in \cref{sec:method}.
Processing a 10s window takes the system 54.35ms, implying that a new heart rate can be predicted by the system every 5s (due to the 5s second overlap between windows). This latency is an adjustable parameter of the system based on the overlap ratio.
%\dm{I think the duration between adjacent HR estimation is still 2s, as the previous HR also takes 63ms to compute, except for the first estimation. So I think we can remove 2.06s as it is also an adjustable parameter depending on the overlapping ratio. }
The system power consumption is given in \cref{tab:power}. 
Overall, the full system (including microphone sampling, denoising and HR prediction) consumes 701mW. The microphone sampling runs continuously, but the hEARt system is only active for 65.04ms for each estimate, and an estimate is made every 5s. Thus, the average energy consumed per second is $(2871-2775)mW\times1s+(3547-2871)mW\times65.04ms/5 = 104.79mJ$. To place this in context, if run on a wireless earbud such as the Apple Airpod Pro (with a battery capacity of 43mAh\footnote{https://www.ifixit.com/Teardown/AirPods+Pro+Teardown/127551}), hEARt could operate continuously for a time $T = \frac{43mAh\times5V}{104.79mJ}=2.05hr$. While this may seem like a short operating time, this system has been implemented on a power hungry Raspberry Pi without optimizing for energy consumption. By implementing the model on a low power microcontroller, power consumption will be reduced. Additionally, when converting the denoising CNN to Tensorflow Lite, optimizations and quantization were not applied. The model can thus be further optimized to reduce both energy consumption and latency. With such optimizations, the energy expenditure will be much lower. However, ultimately this gives an indication of how hEARt could feasibly be implemented on a commercial earbud. 

% \begin{table}[ht]
%     \centering
%      \caption{Latency.}
%      \resizebox{0.6\columnwidth}{!}{%
%     \begin{tabular}{|cc|}
%     \hline
%        Operation (window) & Latency (ms) \\
%        \hline
%        \hline
%        Preprocessing (2s) & 1.66\\
%        Denoising (2s)& 7.66\\
%     Reconstruction (10s) & 17.96\\
%        HR extraction (10s) & 0.48\\
%        \hline
%        Total (10s) & 65.04 \\
%         \hline
%     \end{tabular}
%      }
%      \label{tab:latency}
% \end{table}

% \begin{table}[ht]
%     \centering
%      \caption{Power consumption.}
%      \resizebox{0.6\columnwidth}{!}{%
%     \begin{tabular}{|cc|}
%     \hline
%        Operation & Power (W) \\
%        \hline
%        \hline
%        RasPi (Baseline) & 2.775\\ 
%        RasPi+Mic & 2.871\\
%      Full system & 3.547\\
%         \hline
%     \end{tabular}
%     \label{tab:power}
%     }
% \end{table}

\begin{table}[ht]

\begin{minipage}
{.45\columnwidth}
  
    \centering
     \caption{Latency.}
     \resizebox{1\columnwidth}{!}{%
    \begin{tabular}{|cc|}
    \hline
       Operation (window) & Latency (ms) \\
       \hline
       \hline
       Preprocessing (2s) & 1.66\\
       Denoising (2s)& 7.66\\
    Reconstruction (10s) & 17.96\\
       HR extraction (10s) & 0.48\\
       \hline
       Total (10s) & 65.04 \\
        \hline
    \end{tabular}
     }
     \label{tab:latency}
\end{minipage}
    \hfill
\begin{minipage}{.45\columnwidth}
    \centering
     \caption{Power consumption.}
     \resizebox{1\columnwidth}{!}{%
    \begin{tabular}{|cc|}
    \hline
       Operation & Power (W) \\
       \hline
       \hline
       RasPi (Baseline) & 2.775\\ 
       RasPi+Mic & 2.871\\
     Full system & 3.547\\
        \hline
    \end{tabular}
    \label{tab:power}
    }
\end{minipage}
\end{table}

% \begin{table}[ht]
%     \centering
%      \caption{Power consumption of hEARt.\dm{by rephrasing the text, the two tables can be arranged in the same row (side by side)}.}
%     \begin{tabular}{|cc|}
%     \hline
%        Operation & Power (W) \\
%        \hline
%        \hline
%        RasPi (Baseline) & 2.775\\ 
%        RasPi+Mic & 2.871\\
%      Full system & 3.547\\
%         \hline
%     \end{tabular}
%     \label{tab:power}
% \end{table}

% 96mW*1s+605 mW*63.67ms = 143.52mJ. T = 81mAHh*5V/143.52mW = 3.01hr.

\section{Discussion}
While we acknowledge the merits of PPG-based HR monitoring and are aware of the wealth of information PPG carries, there is great value in showing the potential of a lesser explored modality: in-ear microphones.
% As suggested by industrial research and market trends, ultra-miniaturized form factors will likely have a dominant role, especially as the distinction between hearing aids and earables becomes less marked.
% Additionally, 
In-ear microphones offer substantial advantages over PPG, including their price tag and prevalence in high-end earbuds and hearing aids, due to their importance in adaptive noise cancellation. Microphones are also relatively power efficient sensors~\cite{datasheet:SPU1410LR5H-QB}, requiring less current than PPG (especially when used with high intensity configurations to increase SNR)~\cite{datasheet:MAXM86161}.
Concretely, the microphone we use~\cite{datasheet:SPU1410LR5H-QB} has a current draw of $0.12mA$, more than 10 times less than that of a state-of-the-art wearable dedicated PPG module, the MAXM86161, which draws $1.62mA$~\cite{datasheet:MAXM86161}.
%Even using the lower power configuration, the MAXM86161 optical module draws around $1.62mA$. %Notably, if we were after higher PPG SNR, we would use their most performing configuration, which draws $3.78mA$.

In the remainder of this section we reason over some shortcomings of our work, and potential solutions.
First, we are aware of the limitations that come with a simple, cheap prototype like ours.
For instance, some of the collected data was corrupted as the subjects were unable to wear the earbuds properly, even though they were asked to fit them tightly. 
This indicates that proper sealing of the ear canal is critical. Given that people have different shaped and sized ear canals, it is necessary to select the optimal ear tip size for each individual to improve performance, using an automated method of checking the fit of the earbuds and the seal as done in \cite{Ma2021OESense}. % or as that implemented by Apple for the AirPod Pro\footnote{https://support.apple.com/en-us/HT210633}. 
The data corruption while running was also worsened by the wires on the earbuds which move during vigorous activity thus dislodging the earbuds. 
Using a wireless prototype would thus improve earbud fit and resulting system performance.
Interestingly, fit and positioning issues have also been reported for in-ear PPG~\cite{ferlini2021ear}.
Though, contrary to PPG where sensor misplacement can be hard to identify and may lead to artefacts, poor fit is obvious with in-ear microphones~\cite{Ma2021OESense}.
Nonetheless, our work shows the viability of using in-ear microphones for the detection of HR, even with a far-from-optimal prototype.

We note that the MAPE in HR estimation while speaking requires improvement to meet ANSI standards for HR monitors. Since the system aims to determine HR under active conditions (e.g., running), we expect the amount of speaking to be less than in non-active scenarios, limiting the impact of those errors especially over prolonged periods. The errors occur since speaking introduces non-stationary noise that is different than walking and running. Other techniques to remove non-stationary noise can be considered, or the quantity of speaking samples could be increased during training so that the model can learn characteristics of these signals better.

Given that earbuds are mainly used for audio delivery, one concern is whether music playback will affect performance. As studied by Ma et al.~\cite{Ma2021OESense}, in music, the average energy ratio below 50~Hz is only 1.5\%. This means that music has a negligible impact on our system as it operates on signals below 50~Hz. To confirm this, we first superimpose music on the collected in-ear signals and filter it with a lowpass filter (pass band $<$50Hz). We then compute the Pearson correlation between the filtered signal and original signal, yielding a coefficient of 0.982, further proving our system's robustness.

Finally, despite the good performance, more strategies can be utilized for further improvement. Firstly, we expect that fine tuning a model for each activity will improve activity level performance. We also aim to investigate the use of a LSTM-based model to better model dependencies between adjacent HS. Additionally, collecting data from more subjects encompassing a wider range of HR will improve the ability of the model to further generalize to higher HR. 
%An additional aspect left unexplored is whether personalization can help improve HR estimation. To achieve this, models could be fine-tuned using data from an individual user. Due to the differences in HRs and in HS seen in the population, we expect that training the model using some of the user's own data would improve prediction accuracy. This would be particularly relevant for users with poor overall performance, as this indicates that the user's data is significantly different from that of the population. However, additional challenges to be solved around this involve the collection of this type of user data with reasonable labels.

\section{Related Work}

\textbf{Earables:}
Earables have attracted tremendous attention for human sensing applications, especially for health and wellbeing monitoring~\cite{EarablesForPersonalScaleBehaviorAnalytics}. Literature has investigated earables for blood flow and oxygen consumption~\cite{LeBoeuf2014Earbud-BasedVO2max}, 
dietary monitoring and swallow detection~\cite{amft2005analysis}, blood pressure monitoring~\cite{EBPWearableSystemForFrequentAndComfortable}, step counting~\cite{Ma2021OESense}, heart and respiratory rate tracking~\cite{InEarAudioWearableMeasurementOfHeartAnd}, user identification and gesture recognition~\cite{fan2021headfi}, etc.
% Bui et al.~\cite{EBPWearableSystemForFrequentAndComfortable} proposed a device to measure blood pressure from the artery in the ear canal. Amft et al.~\cite{amft2005analysis} developed a monitoring system that classifies different kinds of food by analysing chewing sounds. 
% With respect to motion tracking, facial expression tracking shows great potential, due to the physical deformations generated by facial muscle movements~\cite{matthies2017earfieldsensing,ando2017canalsense}. 
% An acoustic-based in-ear system using the in ear microphone for step counting, activity recognition, and hand-to-face gesture interaction was also investigated ~\cite{Ma2021OESense}. Respiratory rate measuring and biological analysis are also two vital application fields for earables ~\cite{TowardsRespirationRateMonitoringUsingAnInEar,DetectionOfRespiratorySoundsAtTheExternal,ColorSpectrographicRespiratoryMonitoringFromTheExternal}.
A paradigm named HeadFi was proposed to turn the drivers inside existing headphones into a sensor, with its potential validated in four applications~\cite{fan2021headfi}. %: touch based gesture recognition, user identification, HR monitoring, and voice communication without a microphone~\cite{fan2021headfi}. 
Using the HeadFi system, the authors perform HR monitoring in the stationary case and with the addition of body movement caused by taking the headphones on and off. However, they did not study HR monitoring in the presence of full-body motion %(ie. the intense motions that lead to errors in PPG estimation) 
such as running and walking, or speaking. %They also did not assess the performance of their earable based system while the user was speaking. 

\textbf{Heart Rate Monitoring:}
HR is generally measured using electroencephalogram (EEG), ECG or PPG sensors. However, EEG has limited applications out-of-the-clinic and ECG requires a chest strap, making it inconvenient.  %Some smartwatches can capture ECGs, but require the user to close the circuit with their fingers. 
PPG is the standard for HR monitoring in wearables. However, it is highly susceptible to MAs caused by physical activity or body motion~\cite{Ismail2021HeartReview}. \cite{Bent2020InvestigatingSensors} showed that amongst consumer and research grade wrist-worn wearables, the error of HR estimation was 30\% higher during activity than at rest. A particular problem with PPG is the signal crossover effect where the PPG sensors lock onto a periodic signal from motion (such as walking or running), which is mistaken as the heart signal~\cite{Bent2020InvestigatingSensors, navalta2020concurrent} causing measurement errors. 
Recently, \cite{ferlini2021ear} reported a 27.14\%, 29.84\% and 12.52\% error of PPG sensors in earables for walking, running and speaking respectively,
%has investigated the feasibility of PPG sensors in earables for HR estimation under motion artefacts. The best reported performance shows a 27.14\%, 29.84\% and 12.52\% error for walking, running and speaking respectively, 
quantitatively demonstrating the challenges of PPG in HR estimation under motion. 
%Beside, the piezoelectric sensor was also used to measure the pressure variance of the surface of the ear canal by placing it on the throat for heart rate estimation. However, it was shown to be quite sensitive to the user’s motion~\cite{park2015wearable}. 
Acoustic sensors have also been studied for HR measurements. %and breathing rates measurements. %while most of them are placed on the chest that are either bulky or discomfort to wear. 
Chen et al.~\cite{chen2015algorithm} estimated HR from a small acoustic sensor placed at the neck. %, and ~\cite{kusche2015ear} measured in-ear pulse wave velocity using heart signal as reference. % but the sensor was originally designed for respiratory sound resulting in heart sounds being much attenuated and even corrupted with respiratory
%signals.% \cm{so what?}.%and validated it in heart rate estimation and heart sound classification, but this sensor was originally designed for respiratory sound and heart signals which is not the ideal case for signal acquisition. 
%One other study~\cite{kusche2015ear} has investigated acoustics in the ear canal, measuring in-ear pulse wave velocity using heart signal as reference. %and the other detecting respiratory sounds~\cite{DetectionOfRespiratorySoundsAtTheExternal}. 
\cite{InEarAudioWearableMeasurementOfHeartAnd} examines both heart and breathing rates using microphones placed in the ear canal while stationary. %embedded under workers' hearing protection devices~\cite{InEarAudioWearableMeasurementOfHeartAnd} and showed an accurate extraction with a MAE of 4.3~BPM and a MAPE of 5.6\%. % An in-ear microphones records breath sounds at various rhythms and intensities for 20 subjects, and experimental results showed an accurate extraction with the MAE of 4.3 BPM %with a standard deviation of 2.2 BPM.
Artefacts were found due to minor movement of the subject's body even though all recordings are collected with subjects remaining stationary. %Also, failure in detecting heart rate with strong breathing might indicate that a fast breathing signal is highly overlapped with heartbeats. 
\cite{nirjon2012musicalheart} introduces a earphone that is equipped with an in-ear microphone to measure HR and an IMU to measure activity level. %, which is able to play music based on user's real-time HR and activity levels. 
However, the impact of activity-induced vibrations on the in-ear HS is not investigated.
These findings imply the challenges in HR measurement from earables under motion. We have thus presented an approach that aims to tackle these and offer a solution to measuring HR in realistic settings.

\section{Conclusion}
We proposed an approach for accurate HR estimation using audio signals collected in the ear canal, under motion artefacts caused by daily activities (e.g., walking, running, and talking). Specifically, leveraging deep learning, we eliminate the interference of motion artefacts and recreate clean heart signals, from which we are able to determine HR. We designed a prototype and collected data from real subjects to evaluate the system. Experimental results demonstrate that our approach achieves mean absolute errors of 3.02 $\pm$ 2.97~BPM, 8.12 $\pm$ 6.74~BPM, 11.23 $\pm$ 9.20~BPM and 9.39 $\pm$ 6.97~BPM for stationary, walking, running and speaking, respectively, opening the door to new non-invasive and affordable HR monitoring with usable performance for daily activities. We also discussed some potential strategies to further improve the performance in the future.

\newpage
\bibliographystyle{IEEEtran}
% \bibliography{references_mend.bib}
\bibliography{new_ref.bib}

%%
% %% If your work has an appendix, this is the place to put it.
% \appendix


\end{document}